\documentclass[aos,MSNbibl,nameyear,dvips]{arximspdf}
\usepackage{multirow}
\usepackage{graphicx}
\doi{10.1214/14-AOS1229} 
\volume{42}
\issue{4}
\pubyear{2014}
\firstpage{1394}
\lastpage{1425}

\makeatletter

\newcommand{\eqref}[1]{(\ref{#1})}
\newtheorem{lem}{Lemma}
\newtheorem{thmm}{Theorem}
\newtheorem{coro}{Corollary}
\newproclaim{example}{Example}
\newproclaim{defi}{Definition}
\newproclaim{algo}{Algorithm}
\newproclaim{remark}{Remark}
\newcommand{\np}{\mathrm{np}}
\newcommand{\slp}{\mathrm{sp}}
\makeatother

\begin{document}
\begin{frontmatter}

\title{On the construction of nested space-filling designs}
\runtitle{Construction of nested space-filling designs}

\begin{aug}
\author[a]{\fnms{Fasheng}~\snm{Sun}\thanksref{t2}\ead[label=e1]{sunfs359@nenu.edu.cn}},
\author[b]{\fnms{Min-Qian}~\snm{Liu}\corref{}\thanksref{t3}\ead[label=e2]{mqliu@nankai.edu.cn}}
\and
\author[c]{\fnms{Peter~Z.~G.}~\snm{Qian}\thanksref{t4}\ead[label=e3]{peterq@stat.wisc.edu}}

\thankstext{t2}{Supported by NNSF of China Grant 11101074 and Program
for Changjiang Scholars and Innovative Research Team in University.}
\thankstext{t3}{Supported by NNSF of China Grant 11271205, Specialized
Research Fund for the Doctoral
Program of Higher Education of China Grant 20130031110002, and ``131''
Talents Program of Tianjin.}
\thankstext{t4}{Supported by NSF Grants DMS-10-55214 and CMMI 0969616.}

\runauthor{F.~Sun, M.-Q. Liu and P.~Z.~G. Qian}

\affiliation{Northeast Normal University, Nankai University and\break
University of Wisconsin--Madison}

\address[a]{F. Sun\\
Department of Statistics\\
KLAS and School of Mathematics and Statistics\\
Northeast Normal University\\
Changchun 130024\\
China\\
\printead{e1}}

\address[b]{M. Q. Liu\\
LPMC and Institute of Statistics\\
Nankai University\\
Tianjin 300071\\
China\\
\printead{e2}}

\address[c]{P. Z. G. Qian\\
Department of Statistics\\
University of Wisconsin--Madison\\
Madison, Wisconsin 53706\\
USA\\
\printead{e3}}
\end{aug}

\received{\smonth{10} \syear{2013}}
\revised{\smonth{4} \syear{2014}}

%
\begin{abstract}
Nested space-filling designs are nested designs with attractive low-dimensional
stratification. Such designs are gaining popularity in \mbox{statistics},
applied mathematics and engineering. Their applications include
multi-fidelity computer models, stochastic optimization problems,
multi-level fitting of nonparametric functions, and linking parameters.
We propose methods for constructing several new classes of nested
space-filling designs. These methods are based on a new group
projection and other algebraic
techniques. The constructed designs can accommodate a nested structure
with an arbitrary number of layers and are more flexible in run size
than the existing families of nested space-filling designs. As a
byproduct, the proposed methods
can also be used to obtain sliced space-filling designs
that are appealing for conducting computer experiments with both
qualitative and quantitative factors.
\end{abstract}

%

\end{frontmatter}
%
\section{Introduction}\label{sec1}

Computer experiments are widely used in science and engineering [Fang,
Li and Sudjianto (\citeyear{FLS}), \citet{SWN}]. A large computer program can often
be run with multiple fidelities. \citet{Q09}, \citet{QTW}
and \citet{QAW} introduced the concept of \emph{nested
space-filing design} (NSFD) for running computer codes with two levels
of accuracy. A pair of NSFD $L_1 \subset L_2$ are two nested designs
with the small design used for the more accurate but more expensive
code and the large design used for the less accurate but cheaper code.
These designs have following properties:
\begin{description}
\item\emph{Economy}: the number of points in $L_1$ is smaller than the
number of
points in $L_2$;
\item \emph{Nested relationship}: $L_1$ is nested within $L_2$, that
is, $L_1\subset L_2$;
\item \emph{Space-filling}: the points in both $L_1$ and $L_2$ achieve
uniformity in low dimensions.
\end{description}
The nested relationship makes it easier to adjust or calibrate the
differences between
the two sources.

Multi-fidelity simulation modeling has received considerable attention
over the past few years, especially in the computational fluid dynamics
and finite element analysis
communities where simulation costs are very high. For example, a
finite element analysis code can be run with varying numbers of mesh
sizes, resulting in multiple versions with three or more levels of
accuracy. Multi-fidelity simulation modeling is a common practice in
engineering. Examples include \citet{DVD} for simulating a
GlattGPC-1 fluidized-bed unit, \citet{CAK} for an aircraft
design application and \citet{MPS} for
a submarine propulsion system application, among others. Specifically
in \citet{DVD}, they reported a physical experiment and
several associated computer models for predicting the steady-state
thermodynamic operation point of a GlattGPC-1 fluidized-bed unit. One
physical model $ (T_{2, {\mathrm{ exp}}}) $ and three computer models
$(T_{2, 3}, T_{2, 2}, T_{2, 1})$ are considered. Model $T_{2, 3}$,
which includes adjustments for heat losses and inlet airflow, is the
most accurate (i.e., producing the closest response to $T_{2, {\mathrm{
exp}}}$). Model $T_{2, 2}$ includes only the adjustment for heat losses,
thus is the medium accurate. While model $T_{2, 1}$ does not adjust for
heat losses or inlet airflow and is thus the least accurate. For such
experiments, it is desirable to run a multi-layer experiment using
NSFDs with three or more layers, which makes it easier to model the
systematic differences among the models and implies more observations
are taken for less accurate experiments [cf., \citet{HQ}].

However, NSFDs with more than two layers cannot be constructed by
using the methods in \citet{QTW} and \citet{QAW}. The technical reason
is the modulus projection used in \citet{QTW} cannot be extended to covering more than two layers.
To overcome this limitation, we present a new group-to-group
projection, called the subgroup projection, in this paper and then
construct several new classes of NSFDs
that can accommodate nesting with an arbitrary number of layers and are
more flexible in run size than existing designs of this type. The
subgroup projection is based on a new
decomposition of Galois fields. As far as we are aware, it is also new
in algebra and may have other algebraic applications beyond design of
experiments. Some families of NSFDs with more than two layers can be
constructed from $(t, s)$-sequences with an
infinite number of elements [\citet{HQ}]. In contrast, the
proposed construction here is simpler and only involves a finite number
of points. The constructed designs here can be used for multi-level
fitting of nonparametric functions [\citet{FI,F07,HQ2}] and linking parameters in engineering
[\citet{HDHSS}], all of which involve nested designs with more
than two layers.

The proposed constructions also give new families of sliced
space-filling designs (SSFDs) which can be used to conduct computer
experiments with both qualitative and quantitative factors [\citet{QWW,HSNB,ZQZ}]. Such
computer experiments are often encountered in practice, though most
literature on computer experiments assumes that all the input variables
are quantitative. For example, \citet{SCI}
described a data center computer experiment which involves qualitative
factors (such as diffuser location and hot-air return-vent location)
and quantitative factors (such as rack power and diffuser flow rate).
For conducting such an experiment, \citet{QW09} proposed to use
an SSFD, say $S=(S'_1, \ldots, S'_v)'$, with each slice $S_i$ being
associated with a level combination of the qualitative factors.
Here, when collapsed over the qualitative levels, the points of the quantitative
factors achieve attractive stratification and at any qualitative level,
the values of the quantitative factors are spread uniformly in a
low-dimensional space. An SSFD can also be used to run a computer model
in batches and conduct multiple computer models [\citet{Q12,WMS}]. Note that the subfield projection used in
\citet{QW09} for constructing SSFDs is a special case of the
subgroup projection proposed in this paper, thus more SSFDs can be
constructed here. Moreover, the SSFDs presented in this paper can be
used to conduct computer experiments with asymmetric qualitative factors.

This paper is organized as follows. Section~\ref{sec2} presents some
useful definitions and notation. Section~\ref{sec3} introduces a
decomposition method of Galois fields and a new algebraic projection,
which play a critical role in the proposed construction methods.
Sections~\ref{sec4}--\ref{sec6} provide new methods for constructing
nested orthogonal arrays, sliced orthogonal arrays and nested
difference matrices, along with illustrative examples. Procedures for
generating NSFDs from nested orthogonal arrays and SSFDs from
sliced orthogonal arrays are presented in Section~\ref{sec7}.
Comparisons with existing work and concluding remarks are given in
Section~\ref{sec8}.

\section{Definitions and notation}
\label{sec2}

\emph{Latin hypercube and orthogonal array-based Latin hypercube}.
A \emph{Latin hypercube} $L= (l_{ij})$ with $n$ runs and $m$ factors is
an $n\times m$ matrix in which each column is a permutation of $0,
\ldots, n-1$ [\citet{MBC}]. Let $A$ be an
orthogonal array $\mathit{OA}(n, m, s, t)$ with levels $0, \ldots, s-1$
[\citet{HSS}]. If we replace the $q=n/s$ zeros in
each column of $A$ by a permutation of $0, \ldots, q-1$, replace the
$q$ ones by a permutation of $q,
\ldots, 2q-1$, and so on, we obtain an \emph{orthogonal array
(OA)-based Latin hypercube}
that achieves stratification up to $t$ dimensions [\citet{T93}].

\emph{Sliced orthogonal array}.
Let $A$ be an $\mathit{OA}(n_2, m, s_2, t)$. Suppose that the rows of $A$ can be
partitioned
into $v$ subarrays of $n_1$ rows, denoted by $A_1, \ldots, A_v$.
Further suppose
that there is a projection $\rho$ that collapses the $s_2$ levels of
$A$ into $s_1$ levels with $s_2>s_1$ and $A_i$ becomes an $\mathit{OA}(n_1, m,
s_1, t)$ after level-collapsing according to $\rho$. Then $A$, or more
precisely $(A_1, \ldots, A_v; \rho)$, is a \emph{sliced orthogonal
array} (SOA) [\citet{QW09}].

\emph{Nested orthogonal array and nested difference matrix}. \citet{QTW} and \citet{QAW} introduced the definition
of nested orthogonal array with two layers, we now extend the
definition to a more general case. Suppose $A_I$ is an $\mathit{OA}(n_I, m, s_I,
t)$ and $\rho_j$ for $j=1, \ldots, I$ are a series of projections
satisfying that $\rho_i(\alpha)=\rho_i(\beta)$ implies $\rho_j(\alpha
)=\rho_j(\beta)$ for $j\leq i$.
Then $(A_1, \ldots, A_I; \rho_1$, $\ldots, \rho_I)$ is\vadjust{\goodbreak} called a \emph
{nested orthogonal array} (NOA) with $I$ layers, denoted by $\mathit{NOA}(A_1$,
$\ldots, A_I; \rho_1, \ldots, \rho_I)$, if:
\begin{longlist}[(ii)]
\item[(i)] $A_{i-1}$ is nested within $A_i$ for $2\leq i\leq I$, that is,
$A_1\subset A_2\subset\cdots\subset A_I$;
\item[(ii)] $\rho_j(A_i)$ is an $\mathit{OA}(n_i, m, s_j, t) \mbox{ for } j\leq i$,
\end{longlist}
where $n_1<n_2<\cdots<n_I$ and $s_1<s_2<\cdots<s_I$. Given a difference
matrix $D(r_I, c, s_I)$ [\citet{BB2}], the concept of nested
difference matrix (NDM) with $I$ layers, denoted by $\mathit{NDM}(D_1,
 \ldots,D_I; \rho_1, \ldots, \rho_I)$, is defined in a similar fashion.

\begin{figure}

\includegraphics{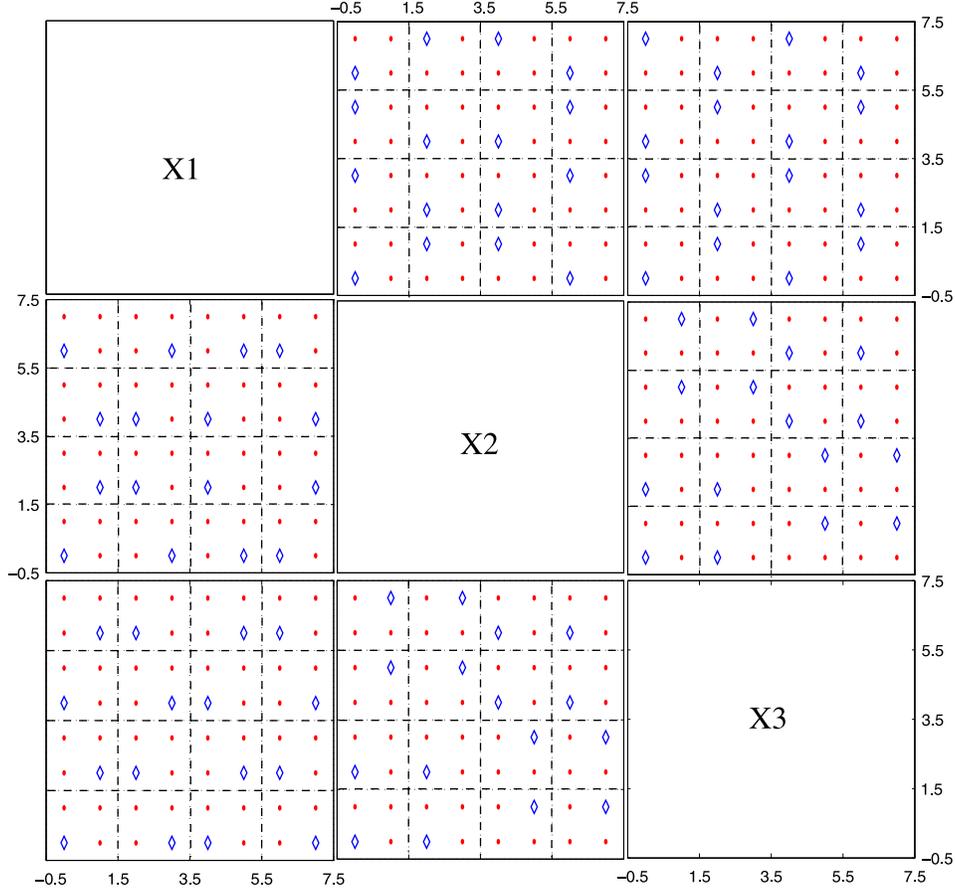}

\caption{Bivariate projections of $A_1$ and $A_2$ with $A_1\subset
A_2$, where the points labeled with both ``$\diamond$'' and
``$\cdot$'' correspond to $A_2$, and those labeled with ``$\diamond$''
correspond to $A_1$.} \label{newqf}
\end{figure}

Note that the concept of NOA here is different from the one introduced
in Mukerjee, Qian and Wu (\citeyear{MQW}), since the $A_i$ for $i=1, \ldots, I-1$
here are not necessarily OAs before the level-collapsing but can still
achieve stratification on any two dimensions. This makes the
construction more flexible. For example, Figure~\ref{newqf}\vadjust{\goodbreak} presents
the bivariate projections of an $\mathit{OA}(64, 5, 8, 2)$ with levels $0, \ldots, 7$, denoted by $A_2$, and a 16-run subset of $A_2$, denoted by $A_1$,
where the points labeled with both
``$\diamond$'' and ``$\cdot$'' correspond to $A_2$, and those labeled
with ``$\diamond$''
correspond to $A_1$ (for saving space, only the bivariate projections
of the first three dimensions are presented here). Obviously, $A_1$ is
not an OA, but it becomes an $\mathit{OA}(16, 5, 4, 2)$ with levels $0, 2, 4, 6$
after the level-collapsing according to the projection $\{0, 1\}
\rightarrow0, \{2, 3\}\rightarrow2, \{4, 5\}\rightarrow4, \{6, 7\}
\rightarrow6$, and the points of $A_1$ achieve stratification on the
$4\times4$ grids in any two dimensions. According to Theorem~1 of
Mukerjee, Qian and Wu (\citeyear{MQW}), if an $\mathit{OA}(N, 5, 8, 2)$ contains an
$\mathit{OA}(16, 5, 4, 2)$, then $N$ must satisfy $N\geq96$, but here the larger
{OA} only has $N=64$ runs if the projection is used to get the smaller OA
with 16 runs. Thus, in the present paper, suitable projections are
critical for the definition and construction of NOAs, and the use of
projections makes the construction more flexible.

Consider two matrices $A=(a_{ij})=(a_1, \ldots, a_s)$ of order $r\times
s$ and $B=(b_{ij})=(b_1, \ldots, b_v)$ of order $u\times v$,
respectively. Their \emph{Kronecker sum} is an $ru\times sv$ matrix
%
\begin{equation}
\label{krosum} A\oplus B=(a_{ij}J+B) \qquad\mbox{where } J \mbox{ is the } u
\times v \mbox { matrix of ones.}
\end{equation}
For $s=v$, here we introduce an operation called \emph{column-wise
Kronecker sum} of $A$ and $B$, given as
%
\begin{equation}
\label{ckrosum} A\oplus_c B= ( a_{1}\oplus b_1,
\ldots, a_{s}\oplus b_s ),
\end{equation}
where $\oplus$ is defined in \eqref{krosum}. These two operations will
be used to
construct NOAs, SOAs and NDMs in the following sections.

\emph{Generator matrix and Rao--Hamming construction}.
Let $s=p^{u}$, $\mathit{GF}(p)\subseteq F_1\subseteq \mathit{GF}(s)$ with $|F_1|=m$,
where $p$ is a prime number and $|F_1|$ denotes the cardinality of set
$F_1$, and let $z_j$ be a column vector of length $k$ with the $j$th
component being one and all
the others being zero, $j = 1, \ldots, k$. We then obtain a $k\times
(m^{k}-1)/(m-1)$ matrix $Z_1$ by collecting all the nonzero column
vectors given by
%
\begin{equation}
\label{gm} z=c_1z_1+ \cdots+c_kz_k\qquad
\mbox{where } c_j\in F_1
\end{equation}
and the first nonzero entry in $(c_1, \ldots, c_k)$
is one. We call $Z_1$ a \emph{generator matrix} over $F_1$
with $k$ independent columns. Let $Z$ be the generator matrix over $\mathit{GF}(s)$
with $k$ independent columns and take all linear combinations of the
row vectors of~$Z$ with coefficients from $\mathit{GF}(s)$, we then obtain an
$\mathit{OA}(s^{k}, (s^{k}-1)/(s-1), s, 2)$. This construction is called the
\emph{Rao--Hamming construction} [\citet{HSS},
Chapter~3].

Lemma~\ref{rh} follows from the Rao--Hamming construction.
%
\begin{lem}\label{rh}
Let $s$ be a prime power and let $A$ be an $s^{k}\times
k$ matrix whose rows consist of all the vectors $(x_1, \ldots,
x_k)$, $x_i\in \mathit{GF}(s), i=1, \ldots, k$, then $AZ$ is an $\mathit{OA}(s^k,
(s^{k}-1)/(s-1), s, 2)$, where $Z$ is a generator matrix over $\mathit{GF}(s)$
with $k$ independent columns.
\end{lem}

\section{A new subgroup projection}\label{sec3}

We now introduce a new projection which will play a key role in the
proposed construction methods in the subsequent sections. Moreover,
this new projection may have other applications in Algebra. We first
present a lemma about the decomposition of Galois fields.

\subsection{Decomposition of Galois fields}

For a finite set $A$ of size $|A|$, put its elements in an column vector
$V_{A}$ with zero being placed as the first entry if included. The
following lemma paves the way for a new decomposition of Galois fields.

\begin{lem}\label{mlem1}
Suppose that $G$ is a finite Abelian group with $|G|=n$. Then there
exists a decomposition of $n=p^{t_1}_1\times\cdots\times p^{t_l}_l$ and cyclic
groups $G_i$ with $|G_i|=p^{t_i}_i$ satisfying
$V_{G}=V_{G_1}\oplus\cdots\oplus V_{G_l}$, where $p_i$ is a prime,
$G_i\subset G$ and $G_i\cap G_j=\{0\}$ for $i\neq j, i, j=1,
\ldots, l$.
\end{lem}

This lemma is a direct result of the fundamental theorem of finite
Abelian group which states that any finite Abelian group can be
decomposed as a direct sum of cyclic subgroups of prime power order
[cf. \citet{H96}, Theorem~2.10.3]. Based on Lemma~\ref{mlem1}, we
have the following result.

\begin{lem}\label{peiji}
Suppose $F_3$ is a Galois field $\mathit{GF}(p^{u_3})$ and
$F_1, F_2$ are subgroups of $F_3$ under operation ``$+$''.
If $F_1$ is a subgroup of $F_2$ under operation ``$+$'', then
there exists a subgroup $T$ of $F_2$ under operation ``$+$''
satisfying $V_{F_2}=V_{F_1} \oplus V_{T}$.
\end{lem}

\begin{pf} Suppose $|F_2|=p^{u_2}$. By Lemma~\ref{mlem1}, there
exists a decomposition of $p^{u_2}=p^{t_1}\times\cdots
\times p^{t_l}$ and cyclic groups $G_i$ satisfying
$V_{F_2}=V_{G_1}\oplus\cdots\oplus V_{G_l}$, where
$|G_i|=p^{t_i}$, $G_i\subset F_2$ and $G_i\cap G_j=\{0\}$ for $i, j=1,
\ldots, l, i\neq
j$. Since the characteristic of
$F_3$ is the prime number $p$, $l=u_2$ and $t_i=1$ for $i=1, \ldots,
l$. That is, $V_{F_2}=V_{G_1}\oplus\cdots\oplus V_{G_{u_2}}$, and
$|G_i|=p, i=1, \ldots, u_2$. As $F_1$ is a subgroup of $F_2$
under operation ``$+$'', without loss of generality, write
$V_{F_1}=V_{G_1}\oplus\cdots\oplus V_{G_{u_1}}$, where $ u_1<u_2$.
Let $V_{T}=V_{G_{u_1+1}}\oplus\cdots\oplus V_{G_{u_2}}$, where
$T$ is a subgroup of $F_2$ under operation ``$+$'', and
$V_{F_2}=V_{F_1} \oplus V_{T}$.
\end{pf}

We now introduce a new decomposition of Galois fields, serving as a
basis for a new group projection. Unless otherwise specified, assume hereinafter
$F_I=\mathit{GF}(s_{I})$, $F_{i-1}$ is a subgroup of $F_{i}$ under operation ``$+$'' for
$i=2, \ldots, I$, and $F_i$ has $s_i=p^{u_i}$ elements for $i=1, \ldots, I$. Then by
Lemma~\ref{peiji}, there exist $T_j$'s satisfying that
%
\begin{equation}
\label{peiji1} V_{F_{i}}=V_{T_1}\oplus V_{T_2}\oplus
\cdots\oplus V_{T_{i}},\qquad i=1, \ldots, I,
\end{equation}
where $T_1=F_1$ and $T_j$ is a subgroup of $F_j$ for $j=2, \ldots, I$.

We introduce Algorithm \ref{algo1} to perform the decomposition in \eqref{peiji1}.
%
\begin{algo}\label{algo1}
\textit{Step} 1. From $F_1$, obtain
%
\begin{equation}
\label{vt1} V_{T_1}=V_{F_1},
\end{equation}
where the first entry of $V_{T_1}$ is zero.

\textit{Step} 2. For $i=2,\ldots, I$, from $F_{i-1}\subset F_{i}$ and
Lemma~\ref{peiji}, obtain $T_{i}$ as a subgroup of $F_{i}$ under
operation ``$+$'' such that the direct sum of $F_{i-1}$ and $T_{i}$ is
$F_{i}$. That is,
%
\begin{equation}
\label{dsum} V_{F_{i}}= V_{F_{i-1}}\oplus V_{T_{i}}\qquad \mbox{for } i=2,\ldots, I.
\end{equation}

\textit{Step} 3. Combining \eqref{vt1} and \eqref{dsum} gives the
decomposition in \eqref{peiji1}.
\end{algo}

\subsection{A new subgroup projection}
\label{sec32}

Using the above decomposition, we are now ready to propose a new
group-to-group projection, which will play a key role in our
construction of NSFDs. As far as we are aware, this projection is new
in algebra and may have applications in other algebraic problems.

In \eqref{peiji1}, any $\gamma\in
F_{I}$ can be uniquely expressed as
%
\begin{equation}
\label{pei} \gamma=\beta_1+\cdots+\beta_{I}, \qquad
\beta_i\in T_{i} \mbox{ for } i=1, \ldots, I.
\end{equation}
Using \eqref{peiji1} and \eqref{pei}, define a projection $\rho_i\dvtx F_{I}\rightarrow F_i$ as
%
\begin{equation}
\label{eq3} \rho_i(\gamma)= \rho_i(
\beta_1+\cdots+\beta_{I})=\beta_1+\cdots+
\beta_{i},
\end{equation}
which maps an element in $F_I$ to its counterpart in the subgroup
$F_i$, $i=1, \ldots, I$.
We call this projection the \emph{subgroup projection.}

\begin{lem}\label{projection}
For the subgroup projection and $\gamma_1, \gamma_2, \gamma\in F_{I}$,
we have:
\begin{longlist}[{(iii)}]
\item [{(i)}] $\rho_i(\gamma_1+\gamma_2)=\rho_i(\gamma_1)+ \rho
_i(\gamma_2)$;
\item[{(ii)}] $\rho_{i}(\rho_{j}(\gamma))=\rho_{\min\{i, j\}
}(\gamma)\in F_{\min\{i, j\}}$;
\item[{(iii)}] $\rho_i(\gamma_1)=\rho_i(\gamma_2)$ implies $\rho
_j(\gamma_1)=\rho_j(\gamma_2)$ for $ j\leq i$;

\item[{(iv)}] $\rho_{i}(V_{F_{I}})=V_{F_{i}}\otimes{\mathbf
1}_{s_{I}/s_i}$,
\end{longlist}
where ${\mathbf 1}_n$ denotes the $n$th unity vector.
\end{lem}

Lemma~\ref{projection2} gives some desirable properties of the
subgroup projection.
%
\begin{lem}\label{projection2}
\textup{(i)} If $D$ is a $D(r, c, s_{{i}})$ based on $F_i$,
then $\rho_{j}(D)=(\rho_{j}(d_{uv}))$ is a $D(r, c, s_{j})$ based on $F_j$
for $1\leq j\leq i\leq I$.
\begin{longlist}[(ii)]
\item[{(ii)}] If $A$ is an $\mathit{OA}(n, m, s_{{i}}, t)$ based on $F_i$,
then $\rho_{j}(A)=(\rho_{j}(a_{uv}))$ is an $\mathit{OA}(n, m, s_{{j}},
t)$ based on $F_j$ for $1\leq j\leq i\leq I$.
\end{longlist}
\end{lem}

The subgroup projection works under a \emph{subgroup} structure and is
more general than the subfield projection introduced in \citet{QW09} and the modulus projection in
\citet{QTW}. The modulus projection, denote by $\varphi$,
satisfies Lemma~\ref{projection2}, but does not satisfy Lemma~\ref
{projection}. Thus, the method in \citet{QTW} cannot be
extended to construct NSFDs with more than two layers. For
illustration, take $F_1=\mathit{GF}(2)$, $F_2=\mathit{GF}(2^2)$ and $F_3=\mathit{GF}(2^3)$ with
irreducible polynomials $g_1(x)=x+1$, $g_2(x)=x^2+x+1$ and
$g_3(x)=x^3+x+1$, respectively. For any $f(x)\in F_3$, $\varphi$ gives
\[
\varphi_3\bigl(f(x)\bigr)=f(x), \qquad \varphi_2\bigl(f(x)
\bigr)=f_{g_2(x)}(x),\qquad \varphi _1\bigl(f(x)\bigr)=f_{g_1(x)}(x),
\]
where $f_{g(x)}(x)$ denotes the residue of $f(x)$ modulo $g(x)$. Here,
$\varphi_2(x^2)=\varphi_2(x+1)=x+1$,
but $\varphi_1(x^2)=1\neq0=\varphi_1(x+1)$, which implies $\varphi$
does not satisfy Lemma~\ref{projection}. The truncation projection used
in \citet{QAW} for constructing NDMs satisfies Lemmas \ref
{projection} and \ref{projection2} and is a special form of the
subgroup projection.

The subgroup projection will be extended to a more general group
structure in Section~\ref{sec6}.

\section{Construction of NOAs and SOAs using the Rao--Hamming method for
the case of $u_i<u_{i+1}$}\label{sec4}

We now present new methods to construct NOAs with two or more layers
and a sliced structure. Suppose $F_I=\mathit{GF}(s_{I})$, $F_i=\{f(x)\in F_I|
\mbox{ the degree of } f(x)$
is less than or equal to $u_i-1\}$, $s_i=p^{u_i}$, for $i=1, \ldots,
I$, and $u_{i-1}<u_{i}$ for $i=2, \ldots, I$. Then $F_{i-1}$ is a
subgroup of $F_{i}$ under operation ``$+$'' for $i=2, \ldots, I$, and \eqref
{peiji1}, \eqref{pei}
and Lemma~\ref{projection} hold.

\begin{algo}\label{newalgo1}
\textit{Step} 1. Let $G_i= {F_i\times\cdots\times
F_i} =\{(\gamma_1, \ldots, \gamma_k)|\gamma_j\in F_i, j=1, \ldots, k\}
$, $i=1, \ldots, I$.
For any elements\vadjust{\goodbreak} $(\gamma_{11}, \ldots, \gamma_{1k})$ and $(\gamma
_{21}, \ldots,
\gamma_{2k})\in G_i$, define
$(\gamma_{11}, \ldots, \gamma_{1k})+(\gamma_{21}, \ldots, \gamma
_{2k})=(\gamma_{11}+\gamma_{21}, \ldots,
\gamma_{1k}+\gamma_{2k})$, where the operation ``$+$'' is the addition on
$F_i$.

\textit{Step} 2. Let
$W_i=\{(\gamma_1, \ldots, \gamma_k)|\gamma_j\in
T_i, j=1, \ldots, k\}$,
which can be expressed as
$\{{\mathbf 0}'_k, {\bolds\beta^i_1 },
\ldots, {\bolds\beta^i_{(s_{i}/s_{i-1})^k-1}}\},   i=1, \ldots, I$,
where ${\mathbf 0}_k$ is the $k$th zero vector and $s_0=1$.

\textit{Step} 3. Suppose $G_1=\{{\mathbf 0}'_k, {\bolds\eta}_1,
\ldots, {\bolds\eta}_{s^k_1-1}\}$. Define an $s^k_1\times k$
matrix $H_1$ to be
$H_1=({\mathbf 0}_k, {\bolds\eta}'_1, \ldots, {\bolds\eta}'_{s^k_1-1})'$.
For $i=2, \ldots, I$, let
%
\begin{equation}\quad
\label{peiji2} H_{i}=\bigl(H'_{i-1},\bigl[{
\bolds\beta^{i}_1}\oplus_c H_{i-1}
\bigr]',\ldots, \bigl[{\bolds\beta^{i}_{(s_{i}/s_{i-1})^k-1}}
\oplus_c H_{i-1}\bigr]'\bigr)',\qquad i=2,\ldots,I,\hspace*{-4pt}
\end{equation}
where $\oplus_c$ is defined in \eqref{ckrosum}. Obtain
%
\begin{equation}\qquad
\label{peiji3} H_{I}=\bigl(H'_i,\bigl[{
\bolds\alpha^{i}_1} \oplus_c H_i
\bigr]', \ldots, \bigl[{\bolds\alpha^{i}_{(s_{I}/s_i)^k-1}}
\oplus_c H_i\bigr]'\bigr)', \qquad i=1, \ldots, I-1,
\end{equation}
where ${\bolds\alpha^i_j}=(\alpha^i_{j1}, \ldots, \alpha^i_{jk})
\in G_{I}\setminus G_{i}$ for $j=1, \ldots, (s_{I}/s_i)^k-1$.\vspace*{1pt}

\textit{Step 4.} Let
\begin{eqnarray*}
A_i &=& H_iC\qquad \mbox{for } i=1,
\ldots, I,
\\
{\bolds\gamma}^i_j&=&{\bolds\alpha}^i_jC\qquad
\mbox{for } i=1, \ldots, I-1, j=1, \ldots, (s_{I}/s_i)^k-1,
\\
{\bolds\delta}^{i}_j&=&{\bolds\beta}^{i}_jC\qquad
\mbox{for } i=2, \ldots, I,  j=1, \ldots, (s_{i}/s_{i-1})^k-1,
\mbox{ and}
\\
\Gamma^i_l&=&A_I\bigl(\bigl[(l-1)s_i^k+1
\bigr]\dvtx ls_i^k\bigr)\qquad \mbox{for } i=1, \ldots, I-1, l=1, \ldots, (s_I/s_i)^k,
\end{eqnarray*}
where $C$ is a generator matrix over $\mathit{GF}(p)$ with $k$ independent columns,
and for any matrix $A$, $A(u\dvtx v)$ denotes its submatrix consisting of
rows $u$ to $v$.
\end{algo}

\begin{thmm}\label{mthm1}
For the $A_i$'s and $\Gamma^i_l$'s constructed in Algorithm \ref
{newalgo1}, and $\rho_i$'s defined in Section~\ref{sec32}, we have:
\begin{longlist}[{(iii)}]
\item[{(i)}]
$ A_{I}=(A'_i, ({\bolds\gamma}^i_1 \oplus_c A_i)', \ldots,
({\bolds\gamma}^i_{(s_{I}/s_i)^k-1} \oplus_c A_i)')', \mbox{ for }
i=1, \ldots, I-1$,\ $ A_{i}=(A'_{i-1}, ({\bolds\delta}^{i}_1 \oplus_c A_{i-1})',
\ldots,
({\bolds\delta}^{i}_{(s_{i}/s_{i-1})^k-1} \oplus_c A_{i-1})')',
\mbox{ for } i=2, \ldots, I$.

\item[{(ii)}] $(A_1, \ldots, A_I; \rho_1, \ldots, \rho_I)$ is an
NOA with $I$ layers, where $\rho_j(A_i)$ is an $\mathit{OA}(s^k_i,
(p^k-1)/(p-1), s_j, 2)$, for $1\leq j\leq i\leq I$;

\item[{(iii)}] $(\Gamma^i_1, \ldots, \Gamma^i_{(s_I/s_i)^k}; \rho
_j)$ is an SOA,
for $1\leq j\leq i\leq I-1$.
\end{longlist}
\end{thmm}

\begin{pf}
{(i)} It follows from the expressions of $H_i$'s in \eqref{peiji2}
and \eqref{peiji3}, and the definition of $A_i$.

{(ii)} From Lemmas \ref{rh} and \ref{projection2}, $\rho_j(A_i)$
is an $\mathit{OA}(s^k_i, (p^k-1)/(p-1), s_j, 2)$ for $j\leq i$, and thus $(A_1,
\ldots, A_I; \rho_1, \ldots, \rho_I)$ is an NOA with $I$ layers;

{(iii)}
Since $\rho_j({\bolds\gamma}^i_l \oplus_c A_i)=\rho_j({\bolds\gamma}^i_l) \oplus_c \rho_j(A_i)$, then $\rho_j({\bolds\gamma
}^i_l \oplus_c A_i)$ is an
$\mathit{OA}(s^k_i,\break  (p^k-1)/(p-1), s_j, 2)$ that can be obtained by permuting
the levels of each factor in $\rho_j(A_i)$. Note that $\Gamma^i_1=A_i$
and $\Gamma^i_l={\bolds\gamma}^i_{l-1} \oplus_c A_i$ for $l>1$,
and thus $(\Gamma^i_1, \ldots, \Gamma^i_{(s_I/s_i)^k}; \rho_j)$ is an
SOA, for $1\leq j\leq i\leq I-1$.
\end{pf}

\begin{remark} \label{rmk3}
If $k>2$ in Theorem~\ref{mthm1}, we can choose some columns from the
generator matrix $C$ to form a new matrix $C^*$ such that the strength
$t$ of $A_I=H_IC^*$ is greater than 2. For $k=3$ and $p=2$, if we take
\[
C^*=\left( \matrix{ 1 & 0 & 0 & 1
\vspace*{2pt}\cr
0 & 1 & 0 & 1
\vspace*{2pt}\cr
0 & 0 & 1 & 1}
 \right)\qquad \mbox{from } C=\left(
\matrix{1 & 0 & 0 & 1 & 0 & 1 &1
\vspace*{2pt}\cr
0 & 1 & 0 & 1 & 1 & 0 &1
\vspace*{2pt}\cr
0 & 0 & 1 & 0 & 1 & 1 &1 }
 \right),
\]
then $A_I=H_IC^*$ has strength 3. Based on such $C^*$'s and $A_I$'s,
the NSFDs and SSFDs
generated in Section~\ref{sec7} will achieve stratification up to $t>2$
dimensions.
\end{remark}

\begin{example}\label{me2}
Let $s_1=2, s_2=2^{2}, s_3=2^{3}$,
$F_1=\{0, 1\}, F_2=\{0, 1, x,
x+1\} \mbox{ and } F_3=\mathit{GF}(2^3)=\{0, 1, x, x+1, x^2, x^2+1, x^2+x,
x^2+x+1\}$.
Here, $F_i$ is a subgroup of $F_{i+1}$ under the operation ``$+$'', $i=1,
2$. From
\eqref{peiji1},
\[
\cases{ %
 V_{F_2}=V_{T_1}\oplus
V_{T_2},
\vspace*{2pt}\cr
V_{F_3}=V_{T_1}\oplus V_{T_2}\oplus
V_{T_3}, }
\]
with $V_{T_1}=(0, 1)', V_{T_2}=(0,
x)'$ and $V_{T_3}=(0, x^2)'$.
For $k=2$,
\begin{eqnarray*}
W_1&=&\bigl\{(0, 0), (0, 1), (1, 0), (1, 1)\bigr\},
\\
W_2&=&\bigl\{(0, 0), (0, x), (x, 0), (x, x)\bigr\},
\\
W_3&=&\bigl\{(0, 0), \bigl(0, x^2\bigr),
\bigl(x^2, 0\bigr), \bigl(x^2, x^2\bigr)\bigr
\},
\\
H_1&=&\left( %
\matrix{0 & 0
\vspace*{2pt}\cr
0 & 1
\vspace*{2pt}\cr
1 & 0
\vspace*{2pt}\cr
1 & 1 }
 \right),\qquad H_2 = \left(
\matrix{ H_1
\vspace*{2pt}\cr
(0, x) \oplus_c H_1
\vspace*{2pt}\cr
(x, 0) \oplus_c H_1
\vspace*{2pt}\cr
(x, x) \oplus_c H_1}
 \right)\quad \mbox{and}\quad H_3= \left( %
\matrix{
H_2
\vspace*{2pt}\cr
\bigl(0, x^2\bigr) \oplus_c H_2
\vspace*{2pt}\cr
\bigl(x^2, 0\bigr) \oplus_c H_2
\vspace*{2pt}\cr
\bigl(x^2, x^2\bigr) \oplus_c
H_2 }
 \right).
\end{eqnarray*}
Let $C$ be a generator matrix over $\mathit{GF}(2)$ with two independent columns
given by
\[
C=\left( \matrix{ 1 & 0 & 1
\vspace*{2pt}\cr
0 & 1 & 1 }
 \right).
\]
Table~\ref{mtb1} gives $A_1, A_2, A_{3}$ and $\Gamma^i_l$ for $i=1, 2$
and $l=1, \ldots, 4^{3-i}$.
%
\begin{table}
\caption{The matrix $A_3$ in Example \protect\ref{me2}, where
$A_1=A_3(1\dvtx 4)$, $A_2=A_3(1\dvtx 16)$,
$\Gamma^1_l=A_3([4(l-1)$ $+1]\dvtx 4l)$ for $l=1, \ldots, 16$, and
$\Gamma^2_l=A_3([16(l-1)+1]\dvtx 16l)$ for $l=1, \ldots, 4$}
\label{mtb1}
%
\begin{tabular*}{\textwidth}{@{\extracolsep{\fill}}lccccccc@{}}
\hline
\textbf{Row} & $\bolds{x_1}$ & $\bolds{x_2}$ & $\bolds{x_3}$ &\textbf{Row} & $\bolds{x_1}$ & $\bolds{x_2}$ & $\bolds{x_3}$ \\
\hline
\phantom{0}1 & 0& 0& 0 & 33 & $x^2$& 0& $x^2$ \\
\phantom{0}2 & 0& 1& 1 & 34 & $x^2$& 1& $x^2$+1 \\
\phantom{0}3 & 1& 0& 1 & 35 & $x^2$+1& 0& $x^2$+1 \\
\phantom{0}4 & 1& 1& 0 & 36 & $x^2$+1& 1& $x^2$ \\
\phantom{0}5 & 0& $x$& $x$ & 37 & $x^2$& $x$& $x^2$+$x$ \\
\phantom{0}6 & 0& $x$+1& $x$+1 & 38 & $x^2$& $x$+1& $x^2$+$x$+1 \\
\phantom{0}7 & 1& $x$& $x$+1 & 39 & $x^2$+1& $x$& $x^2$+$x$+1 \\
\phantom{0}8 & 1& $x$+1& $x$ & 40 & $x^2$+1& $x$+1& $x^2$+$x$ \\
\phantom{0}9 & $x$& 0& $x$ & 41 & $x^2$+$x$& 0& $x^2$+$x$ \\
10 & $x$& 1& $x$+1 & 42 & $x^2$+$x$& 1& $x^2$+$x$+1 \\
11 & $x$+1& 0& $x$+1 & 43 & $x^2$+$x$+1& 0& $x^2$+$x$+1 \\
12 & $x$+1& 1& $x$ & 44 & $x^2$+$x$+1& 1& $x^2$+$x$ \\
13 & $x$& $x$& 0 & 45 & $x^2$+$x$& $x$& $x^2$ \\
14 & $x$& $x$+1& 1 & 46 & $x^2$+$x$& $x$+1& $x^2$+1 \\
15 & $x$+1& $x$& 1 & 47 & $x^2$+$x$+1& $x$& $x^2$+1 \\
16 & $x$+1& $x$+1& 0 & 48 & $x^2$+$x$+1& $x$+1& $x^2$ \\
17 & 0& $x^2$& $x^2$ & 49 & $x^2$& $x^2$& 0 \\
18 & 0& $x^2$+1& $x^2$+1 & 50 & $x^2$& $x^2$+1& 1 \\
19 & 1& $x^2$& $x^2$+1 & 51 & $x^2$+1& $x^2$& 1 \\
20 & 1& $x^2$+1& $x^2$ & 52 & $x^2$+1& $x^2$+1& 0 \\
21 & 0& $x^2$+$x$& $x^2$+$x$ & 53 & $x^2$& $x^2$+$x$& $x$ \\
22 & 0& $x^2$+$x$+1& $x^2$+$x$+1 & 54 & $x^2$& $x^2$+$x$+1& $x$+1 \\
23 & 1& $x^2$+$x$& $x^2$+$x$+1 & 55 & $x^2$+1& $x^2$+$x$& $x$+1 \\
24 & 1& $x^2$+$x$+1& $x^2$+$x$ & 56 & $x^2$+1& $x^2$+$x$+1& $x$ \\
25 & $x$& $x^2$& $x^2$+$x$ & 57 & $x^2$+$x$& $x^2$& $x$ \\
26 & $x$& $x^2$+1& $x^2$+$x$+1 & 58 & $x^2$+$x$& $x^2$+1& $x$+1 \\
27 & $x$+1& $x^2$& $x^2$+$x$+1 & 59 & $x^2$+$x$+1& $x^2$& $x$+1 \\
28 & $x$+1& $x^2$+1& $x^2$+$x$ & 60 & $x^2$+$x$+1& $x^2$+1& $x$ \\
29 & $x$& $x^2$+$x$& $x^2$ & 61 & $x^2$+$x$& $x^2$+$x$& 0 \\
30 & $x$& $x^2$+$x$+1& $x^2$+1 & 62 & $x^2$+$x$& $x^2$+$x$+1& 1 \\
31 & $x$+1& $x^2$+$x$& $x^2$+1 & 63 & $x^2$+$x$+1& $x^2$+$x$& 1 \\
32 & $x$+1& $x^2$+$x$+1& $x^2$ & 64 & $x^2$+$x$+1& $x^2$+$x$+1& 0 \\
\hline
\end{tabular*}
\end{table}

Suppose that $\rho_1, \rho_2$ and $\rho_3$ are defined in \eqref{eq3}
given by\vspace*{9pt}
%
\begin{center}
\begin{tabular}{@{}lcccccccc@{}}
\hline
\multicolumn{1}{@{}l}{$\bolds{\gamma}$} & \textbf{0} & \textbf{1} & \multicolumn{1}{c}{$\bolds{x}$} &
\multicolumn{1}{c}{$\bolds{x+1}$} & \multicolumn{1}{c}{$\bolds{x^2}$} &
\multicolumn{1}{c}{$\bolds{x^2+1}$} & \multicolumn{1}{c}{$\bolds{x^2+x}$}
& \multicolumn{1}{c@{}}{$\bolds{x^2+x+1}$}
\\
\hline
$\rho_1(\gamma)$ & 0 & 1 & 0 & 1 & 0 & 1 & 0 & 1
\\
$\rho_2(\gamma)$ & 0 & 1 & $x$ & $x+1$ & 0 & 1 & $x$ & $x+1$
\\
$\rho_3(\gamma)$ & 0 & 1 & $x$ & $x+1$ & $x^2$ &
$x^2+1$ & $x^2+x$ & $x^2+x+1$
\\
\hline
\end{tabular}
\end{center}\vspace*{9pt}
Note that:
\begin{longlist}[{(ii)}]
\item[{(i)}] $\rho_j(A_i)$ is an $\mathit{OA}(4^{i}, 3, 2^j, 2)$ for
$1\leq j\leq i\leq3$, and thus $(A_1, A_2, A_3; \rho_1$, $\rho_2, \rho
_3)$ is an NOA with three layers;\vadjust{\goodbreak}
\item[{(ii)}] $\rho_j(\Gamma^i_l)$ is an $\mathit{OA}(4^{i}, 3, 2^j, 2)$, and
thus $(\Gamma^i_1, \ldots, \Gamma^i_{4^{3-i}}; \rho_j)$
is an SOA, where $\Gamma^i_l=A_3([4^i(l-1)+1]\dvtx 4^il)$, for $l=1,\ldots,
4^{3-i}$ and $1\leq j\leq i\leq2$.
\end{longlist}
\end{example}

\section{Construction of NOAs, SOAs and NDMs for the case of
$u_i|u_{i+1}$}\label{sec5}

Now assume $u_i<u_{i+1}$ and $u_i$ is a factor of $u_{i+1}$, that is,
$u_i|u_{i+1}$. \citet{QA} presented some constructions of NOAs
with two layers for this case. Here, we provide new constructions for
NOAs with two or more layers and a sliced structure, which are more
general than those in \citet{QA}.

\subsection{Construction of NOAs and SOAs using the Rao--Hamming and
Bush's methods}\label{sec51}

\begin{thmm} \label{mthm2}
By replacing $\mathit{GF}(p)$ for generating the generator matrix $C$ in Step 4
of Algorithm \ref{newalgo1} with $F_1=\mathit{GF}(s_1)$, we obtain:
\begin{longlist}[(iii)]
\item[{(i)}]
$A_{I} = (A'_i, ({\bolds\gamma}^i_1 \oplus_c A_i)', \ldots,
({\bolds\gamma}^i_{(s_{I}/s_i)^k-1} \oplus_c A_i)')', \mbox{ for }
i=1, \ldots, I-1$,
$A_{i} = (A'_{i-1}, ({\bolds\delta}^{i}_1 \oplus_c A_{i-1})',
\ldots,
({\bolds\delta}^{i}_{(s_{i}/s_{i-1})^k-1} \oplus_c A_{i-1})')',
\mbox{ for } i=2, \ldots, I$;

\item[{(ii)}] $(A_1, \ldots, A_I; \rho_1, \ldots, \rho_I)$ is an
NOA with $I$ layers, where $\rho_j(A_i)$ is an $\mathit{OA}(s^k_i,
(s_1^k-1)/(s_1-1), s_j, 2)$, for $1\leq j\leq i\leq I$;\vadjust{\goodbreak}

\item[{(iii)}] $(\Gamma^i_1, \ldots, \Gamma^i_{(s_I/s_i)^k}; \rho
_j)$ is an SOA,
for $1\leq j\leq i\leq I-1$.
\end{longlist}
\end{thmm}

\begin{remark} \label{rmk4}
Similarly, as discussed in Remark~\ref{rmk3}, if $k>2$ in Theorem~\ref{mthm2},
then we can choose some columns of the generator matrix $C$ to form a
new matrix $C^*$ such that $A_I=H_IC^*$ has a strength greater than 2.
\end{remark}

For $s_1\geq k-1$ and $F_1=\{v_1, \ldots, v_{s_1}\}$, if we replace
the generator matrix $C$ in Theorem~\ref{mthm2} by the following matrix:
%
\begin{equation}
\label{vmat} V=\pmatrix{1&1&\cdots&1& 0\vspace*{2pt}
\cr
v_1&v_2&
\cdots&v_{s_1}& 0\vspace*{2pt}
\cr
v^2_1&v^2_2&
\cdots&v^2_{s_1}& 0\vspace*{2pt}
\cr
\vdots&\vdots&\vdots&
\vdots& \vdots\vspace*{2pt}
\cr
v^{k-2}_1&v^{k-2}_2&
\cdots&v^{k-2}_{s_1}& 0\vspace*{2pt}
\cr
v^{k-1}_1&v^{k-1}_2&
\cdots&v^{k-1}_{s_1}& 1},
\end{equation}
then we can generate new NOAs and SOAs with strength $k$ based on
Bush's method
[\citet{HSS}, Chapter~3]. For most cases, $k>2$,
and the related NSFDs and SSFDs will achieve stratification up to $k>2$
dimensions.

\begin{thmm}\label{bush}
If in Theorem~\ref{mthm2}, $C$ is replaced by the $V$ in \eqref{vmat}, then:
\begin{longlist}[{(iii)}]
\item[{(i)}]
$A_i$ is an $\mathit{OA}(s^k_i, s_1+1, s_i, k)$, for $i=1, \ldots, I$;

\item[{(ii)}]
$A_{I} = (A'_i, ({\bolds\gamma}^i_1 \oplus_c A_i)', \ldots,
({\bolds\gamma}^i_{(s_{I}/s_i)^k-1} \oplus_c A_i)')', \mbox{ for }
i=1, \ldots, I-1$,
$A_{i} = (A'_{i-1}, ({\bolds\delta}^{i}_1 \oplus_c A_{i-1})',
\ldots,
({\bolds\delta}^{i}_{(s_{i}/s_{i-1})^k-1} \oplus_c A_{i-1})')',
\mbox{ for } i=2, \ldots, I$;

\item[{(iii)}] $(A_1, \ldots, A_I; \rho_1, \ldots, \rho_I)$ is an
NOA with $I$ layers, where $\rho_j(A_i)$ is an $\mathit{OA}(s^k_i, s_1+1, s_j,
k)$, for $1\leq j\leq i\leq I$;

\item[{(iv)}] $(\Gamma^i_1, \ldots, \Gamma^i_{(s_I/s_i)^k}; \rho
_j)$ is an SOA, for $1\leq j\leq i\leq I-1$.
\end{longlist}
\end{thmm}

\subsection{Construction of NOAs and SOAs from NDMs}\label{sec52}

We now propose a new approach for constructing NOAs and SOAs from NDMs.
Theorem~\ref{mthm3} follows from Lemmas \ref{projection} and
\ref{projection2}.\vadjust{\goodbreak}

\begin{thmm}\label{mthm3}
Let $A$ be an $\mathit{OA}(n, m, s_{I}, 2)$, and
\begin{eqnarray*}
V&=&V_{T_{I}}\oplus V_{T_{{I-1}}}\oplus\cdots\oplus
V_{T_{1}},\qquad D=V V'_{T_{1}},
\\
\Delta^i_{l}&=&D\bigl(\bigl[(l-1)s_i+1\bigr]
\dvtx ls_i\bigr)\qquad \mbox{for } l=1, \ldots, s_{I}/s_i,
 i=1, \ldots, I-1,
\\
\Delta(i, k)&=&\bigl(\bigl(\Delta^i_{1}
\bigr)', \ldots, \bigl(\Delta^i_{k}
\bigr)'\bigr)'\qquad \mbox{for } k=1, \ldots,
s_{I}/s_i-1, i=1, \ldots, I-1.
\end{eqnarray*}
Then for $1\leq j\leq i\leq I$,
$k=1, \ldots, s_{I}/s_i-1$ and $l=1, \ldots, s_{I}/s_i$, we have:
\begin{longlist}[{(iii)}]
\item[{(i)}]$D$ is a $D(s_{I}, s_1, s_{I})$, $\Delta^i_{1}$ is a
$D(s_{i}, s_1, s_{i})$, and $A\oplus D$ is an $\mathit{OA}(ns_{I}, ms_1$,
$s_{I}, 2)$;

\item[{(ii)}] $\rho_j(\Delta^i_{l})$ is a $D(s_{i}, s_1, s_{j})$
based on $F_j$, $\rho_j(\Delta(i, k))$ is a $D(ks_{i}, s_1, s_{j})$
based on $F_j$,
$(\Delta(i, k)$, $D; \rho_j, \rho_I)$ is an NDM with two layers, and
$(\Delta^1_1, \ldots,$ $\Delta^{I-1}_1, D; \rho_1, \ldots, \rho_I)$ is
an NDM with $I$ layers;

\item[{(iii)}] $\rho_j(A\oplus\Delta^i_{l})$ is an $\mathit{OA}(ns_i,
ms_1, s_j, 2)$,
$(A\oplus\Delta^i_{1}, \ldots, A\oplus\Delta^i_{s_{I}/s_i}; \rho_j)$
is an SOA,
$(A\oplus\Delta(i, k), A\oplus D; \rho_j, \rho_I)$ is an NOA with two
layers, and $(A\oplus\Delta^1_1, \ldots, A\oplus\Delta^{I-1}_1, A\oplus
D; \rho_1, \ldots, \rho_I)$ is an NOA with $I$ layers.
\end{longlist}
\end{thmm}

\section{Construction of NOAs, SOAs and NDMs with more general numbers
of levels}
\label{sec6}

The constructed NOAs, SOAs and NDMs so far have prime power numbers of levels.
We now present constructions with more general numbers of levels by
using the operation column-wise Kronecker sum defined in \eqref{ckrosum}.

Let $\Psi_i=\{\psi_1, \ldots, \psi_{s_i}\}$ be a group with positive
integer $s_i$, and
%
\begin{equation}
\label{Omegai6} \Omega_i=\bigl\{\psi_j
\omega^{i-1}|\psi_j\in\Psi_i, \omega\mbox{ is an
indeterminate, } j=1, \ldots, s_i\bigr\},
\end{equation}
for $i=1, \ldots, I$. For any entries $\psi_{j_1}, \psi_{j_2}\in\Psi
_i$, there exists $\psi_{j_3}\in\Psi_i$ such that $\psi_{j_1}+\psi
_{j_2}=\psi_{j_3}$ and
define
\[
\psi_{j_1}\omega^{i-1}+\psi_{j_2}
\omega^{i-1}=\psi_{j_3}\omega^{i-1},
\]
which implies $\Omega_i$ forms a group. Let, for $i=1, \ldots, I$,
%
\begin{equation}
\label{Fi6} F_i=\bigl\{\psi_{l_0}+\psi_{l_1}
\omega+\cdots+\psi_{l_{i-1}}\omega^{i-1}|\psi _{l_b}\in
\Psi_{b+1}, b=0, \ldots, i-1\bigr\},
\end{equation}
and for any elements $\alpha=\psi_{l_0}+\psi_{l_1}\omega+\cdots+\psi
_{l_{i-1}}\omega^{i-1}
\mbox{ and } \beta=\psi^*_{l_0}+\psi^*_{l_1}\omega+\cdots+\psi
^*_{l_{i-1}}\omega^{i-1}\in F_i$, define
\begin{eqnarray*}
\alpha+\beta&=&\bigl(\psi_{l_0}+\psi_{l_1}\omega+\cdots+\psi
_{l_{i-1}}\omega^{i-1}\bigr) +\bigl(\psi^*_{l_0}+
\psi^*_{l_1}\omega+\cdots+\psi^*_{l_{i-1}}\omega^{i-1}
\bigr)
\\
&=& \bigl(\psi_{l_0}+\psi^*_{l_0}\bigr)+\bigl(
\psi_{l_1}+\psi^*_{l_1}\bigr)\omega+\cdots +\bigl(
\psi_{l_{i-1}}+\psi^*_{l_{i-1}}\bigr)\omega^{i-1}.
\end{eqnarray*}
Then $F_i=\sigma(\bigcup^i_{l=1}\Omega_l)$ is a group. Note that $F_i$ is
a subgroup of $F_{i+1}$ and thus \eqref{peiji1} and \eqref{pei} hold,
where $T_i=\Omega_i$. Now express the projection in \eqref{eq3} as
%
\begin{eqnarray}
\label{proj6} \rho_i(\gamma)&=&\psi_{l_0}+
\psi_{l_1}\omega+\cdots+\psi_{l_{i-1}}\omega ^{i-1},
\nonumber
\\[-8pt]
\\[-8pt]
\nonumber
 \gamma&=&\psi_{l_0}+\psi_{l_1}\omega+\cdots+\psi_{l_{I-1}}
\omega^{I-1}\in F_I.
\end{eqnarray}
Hence, Lemmas \ref{projection} and \ref{projection2} also hold under
this projection.

\subsection{Construction of NOAs and SOAs with more general number of
levels}
\label{sec61}

First, we propose a method for constructing SOAs and NOAs with two
layers via the column-wise Kronecker sum.

\begin{thmm}\label{thml}
Let $A_i$ be an $\mathit{OA}(n_i, m, s_i, t)$ based on $\Omega_i $ for $i=1, 2$.
Let $B=A_2\oplus_cA_1$, and denote $B=(B_1', \ldots, B_{n_2}')'$,
where $B_i=B([(i-1)n_1+1]\dvtx in_1), i=1, \ldots, n_2$. Then:
\begin{longlist}[{(iii)}]
\item [{(i)}] $B$ is an $\mathit{OA}(n_1n_2, m, s_1s_2, t)$ based on
$F_2=\sigma(\Omega_1\cup\Omega_2)$;
\item[{(ii)}] $B$ or $(B_1, \ldots, B_{n_2}; \rho_1)$ is an SOA,
where $\rho_1(B_i)$ is an $\mathit{OA}(n_1, m, s_1, t)$ for $i=1, \ldots, n_2$;
\item[{(iii)}] $(B^l, B; \rho_1, \rho_2)$ is an NOA with two
layers, where $B^l=(B'_1, \ldots, B'_l)'$ and
$\rho_1(B^l)$ is an $\mathit{OA}(ln_1, m, s_1, t)$ for $l=1, \ldots, n_2-1$.
\end{longlist}
\end{thmm}

\begin{pf} Denote $A_1=(a^1_{1}, \ldots, a^1_{m})=(a^1(i, j))$
and $A_2=(a^2_{1}, \ldots, a^2_{m})=(a^2(i, j))$.
\begin{longlist}[(iii)]
\item[(i)] For any $t$ columns $(b_{i_1}, \ldots, b_{i_t})$ of $B$,
$b_{i_j}=a^2_{i_j}\oplus a^1_{i_j}$. Then
for any $t$-tuple $(\alpha_1, \ldots, \alpha_t)$ in these columns,
$\alpha_j=\gamma_j+\beta_j \in F_2$ with $\gamma_j\in\Omega_2, \beta
_j\in\Omega_1$ for $j=1, \ldots, t$.
Since $A_1$ is an $\mathit{OA}(n_1, m, s_1, t)$ and $A_2$ is an $\mathit{OA}(n_2, m,
s_2, t)$,
then $(\beta_1, \ldots, \beta_t)$ occurs $n_1/s^t_1$ times in
$(a^1_{i_1}, \ldots, a^1_{i_t})$,
and $(\gamma_1, \ldots, \gamma_t)$ occurs $n_2/s^t_2$ times in
$(a^2_{i_1}, \ldots, a^2_{i_t})$.
Thus, $(\gamma_1+\beta_1, \ldots, \gamma_t+\beta_t)=(\alpha_1, \ldots,
\alpha_t)$
occurs $ {n_1n_2}/{(s_1s_2)^{t}}$ times in $(b_{i_1}, \ldots,
b_{i_t})$, which implies $B$ is
an $\mathit{OA}(n_1n_2, m, s_1s_2, t)$ based on $F_2$.

\item[(ii)] Note that $B_i=(a^2(i, 1), \ldots, a^2(i, m))\oplus_c A_1$ and
\[
\rho_1(B_i)=\bigl(\rho_1
\bigl(a^2(i, 1)\bigr), \ldots, \rho_1\bigl(a^2(i,
m)\bigr)\bigr)\oplus_c A_1.
\]
Clearly, $\rho_1(B_i)$ is an $\mathit{OA}(n_1, m, s_1, t)$ that can be obtained
by permuting
levels of each factor in $A_1$ and $(B_1, \ldots, B_{n_2}; \rho_1)$ is
an SOA.

\item[(iii)] The result in (ii) implies that $(B^l, B; \rho_1, \rho_2)$ is
an NOA with two layers.
\end{longlist}
\end{pf}

\begin{example}\label{exk1}
Let $Z_s=\{0, \ldots, s-1\}$, $s_1=6$, $s_2=2$, $\Psi_1=Z_6$ and $\Psi
_2=Z_2$, then
$\Omega_1=Z_6$, $\Omega_2=\{0, \omega\}$, $F_1=Z_6$ and $F_2=\{Z_6,
\omega+Z_6\}$.
By \eqref{peiji1} and \eqref{proj6}, $V_{F_2}=V_{\Omega_1}\oplus
V_{\Omega_2}$ and
\begin{center}
\begin{tabular}{@{}lcccccccccccc@{}}
\hline
\multicolumn{1}{@{}l}{$\bolds{\gamma\in F_2}$} & \textbf{0} & \textbf{1} & \textbf{2}& \textbf{3} &\textbf{4} & \textbf{5} &
\multicolumn{1}{c}{$\bolds{\omega}$}& \multicolumn{1}{c}{$\bolds{\omega+1}$} & \multicolumn{1}{c}{$\bolds{\omega+2}$}&
\multicolumn{1}{c}{$\bolds{\omega+3}$} & \multicolumn{1}{c}{$\bolds{\omega+4}$} & \multicolumn{1}{c@{}}{$\bolds{\omega+5}$}
\\
\hline
$\rho_1(\gamma)$ & 0 & 1 & 2& 3 &4 & 5 & 0 & 1 & 2& 3 &4 & 5
\\
$\rho_2(\gamma)$ & 0 & 1 & 2& 3 &4 & 5 & $\omega$& $\omega+1$ & $\omega+2$&
$\omega+3$ & $\omega+4$ & $\omega+5$
\\
\hline      \vspace*{6pt}
\end{tabular}
\end{center}

Let $A_1$ be an $\mathit{OA}(36, 3, 6, 2)$ based on $\Omega_1$ and $A_2$ be an
$\mathit{OA}(4, 3, 2, 2)$ based on $\Omega_2$, which are listed in Table~\ref{lexk1}.
%
\begin{table}
\tabcolsep=0pt
\caption{The arrays $A_1$ and $A_2$ in Example \protect\ref{exk1}}
\label{lexk1}
\begin{tabular*}{\textwidth}{@{\extracolsep{\fill}}lccccccccccccccccccccccccccccccccccccccccc@{}}
\hline
\multicolumn{36}{c}{$\bolds{A'_1}$}& &\multicolumn{4}{c@{}}{$\bolds{A'_2}$} \\
\hline
0& 0& 0 & 0& 0& 0& 1& 1& 1& 1 & 1& 1& 2& 2&2& 2&2&2&3&3&3& 3& 3&
3&4&4&4&4&4& 4& 5& 5& 5& 5& 5 &5 & &0 & 0 & $\omega$ & $\omega$ \\
0& 1& 2 & 3& 4& 5& 0& 1& 2& 3 & 4& 5& 0& 1&2& 3&4&5&0&1&2& 3& 4&
5&0&1&2&3&4& 5& 0& 1& 2& 3& 4 &5 & &0 & $\omega$ & 0 & $\omega$ \\
5& 3& 4 & 1& 2& 0& 3& 2& 1& 0 & 4& 5& 0& 5&2& 4&3&1&2&1&3& 5& 0&
4&1&4&0&3&5& 2& 4& 0& 5& 2& 1 &3 & &0 & $\omega$ & $\omega$ & 0 \\
\hline
\end{tabular*}
\end{table}

Then $B=A_2\oplus_cA_1=(B'_1, \ldots, B'_4)'$ satisfies:
\begin{longlist}[{(iii)}]
\item [{(i)}] $B$ is an $\mathit{OA}(144, 3, 12, 2)$ based on $F_2$;
\item[{(ii)}] $\rho_1(B_i)$ is an $\mathit{OA}(36, 3, 6, 2)$ for
$i=1, \ldots, 4$, that is, $(B_1, \ldots, B_{4}; \rho_1)$ is an SOA;
\item[{(iii)}] $\rho_1(B^l)$ is an $\mathit{OA}(36l, 3, 6, 2)$, that is,
$(B^l, B; \rho_1, \rho_2)$ is an NOA with two layers, where
$B^l=(B'_1, \ldots, B'_l)'$ for $l=1, 2, 3$.
\end{longlist}
\end{example}

Since an $\mathit{OA}(s^2, s+1, s, 2)$ exists for any prime power $s$, Theorem~\ref{thml} gives the following corollary.
%
\begin{coro}\label{corocomparison}
For a prime power $s_1$ and $s_2=s_1^2$, there exists an SOA $(B_1,
\ldots, B_{s_2}; \rho)$,
where $B=(B'_1, \ldots, B'_{s_2})'$ is an $\mathit{OA}(s_2^2, s_1+1, s_2, 2)$
and $\rho(B_j)$ is
an $\mathit{OA}(s_2, s_1+1, s_1, 2)$ for $j=1, \ldots, s_2$.
\end{coro}

\begin{remark}\label{remarkcomparison}  For a prime power $s_1$ and
$s_2=s^2_1$,
\citet{XHQ} constructed a special SOA $(B_1, \ldots,
B_{s_2}; \rho)$ based on \emph{doubly orthogonal Sudoku Latin squares},
where $B=(B'_1, \ldots, B'_{s_2})'$
is an $\mathit{OA}(s^2_2, s_1, s_2, 2)$, $\rho(B_j)$ is an $\mathit{OA}(s_2, s_1, s_1,
2)$ and each $B_j$ has maximum stratification in one-dimension in the
sense there are $s_2$ different levels in each column of $B_j$, for
$j=1, \ldots, s_2$. In contrast, $B_j$ in Corollary~\ref
{corocomparison} does not achieve maximum stratification in
one-dimension, since there are only $s_1$ different levels in each
column. But the SOAs obtained here have one more column compared with
that of~\citet{XHQ}. In addition, more SOAs can be
constructed through Theorem~\ref{thml} for general $s_1$ and $s_2$.
\end{remark}

Next, we generalize Theorem~\ref{thml} to construct SOAs and NOAs with
more than two layers.
%
\begin{coro}\label{coroc2}
Let $A_{i}$ be an $\mathit{OA}(n_i, m, s_i, t)$ based on $\Omega_i$ and
$B_{i}=A_{i}\oplus_c\cdots\oplus_c A_{1}$ for $i=1, \ldots, I$.
Suppose $\Gamma^i_l=B_I([(l-1)n_1\cdots n_i+1]\dvtx ln_1\cdots n_i)$ for
$l=1, \ldots, n_{i+1}\cdots n_{I}$ and $i=1, \ldots, I-1$. Then:
\begin{longlist}[{(ii)}]
\item [{(i)}] $(B_1, \ldots, B_I; \rho_1, \ldots, \rho_I)$ is an
NOA with $I$ layers, where $\rho_j(B_i)$ is an $\mathit{OA}(n_1\cdots n_i, m$,
$\prod^j_{l=1}s_l, t)$
for $1\leq j\leq i\leq I$;
\item[{(ii)}] $(\Gamma^i_1, \ldots, \Gamma^i_{n_{i+1}\cdots n_I};
\rho_j)$
is an SOA for $1\leq j\leq i\leq I-1$, where $\rho_j(\Gamma^i_l)$ is an
$\mathit{OA}(n_1\cdots n_i, m$, $\prod^j_{l=1}s_l, t)$ for $l=1, \ldots,
n_{i+1}\cdots n_I$.
\end{longlist}
\end{coro}

\subsection{Construction of NDMs with more general numbers of
levels}
\label{sec62}

We present a method for constructing NDMs via the column-wise Kronecker sum.
Similar to Corollary~\ref{coroc2}, we have the following result.
%
\begin{thmm}\label{thmcnd}
Let $D_i$ be a $D(r_i, c, s_i)$ based on $\Omega_i$ and
$E_i=D_i\oplus_c\cdots\oplus_c D_1$ for $i=1, \ldots, I$. Suppose
\[
\Delta^i_l=E_I\bigl(\bigl[(l-1)r_1
\cdots r_i+1\bigr]\dvtx lr_1\cdots r_i\bigr),
\]
for $l=1, \ldots, r_{i+1}\cdots r_I$ and $i=1, \ldots, I-1$. Then:
\begin{longlist}[{(ii)}]
\item[{(i)}] $(E_1, \ldots, E_I; \rho_1, \ldots, \rho_I)$ is an
NDM with $I$ layers,
where $\rho_j(E_i)$ is a $D(r_1\cdots r_i, c, \prod^j_{l=1}s_l)$ for
$1\leq j\leq i\leq I$;
\item[{(ii)}] $\rho_j(\Delta^i_l)$ is a
$D(r_1\cdots r_i, c, \prod^j_{l=1}s_l)$ for $1\leq j\leq i\leq I-1$ and
$l=1, \ldots,\break  n_{i+1}\cdots n_I$.
\end{longlist}
\end{thmm}

\begin{example}\label{exc3}
Let $Z_s=\{0, \ldots, s-1\}$, $s_1=4$, $s_2=3$, $s_3=2$, $\Psi_1=\mathit{GF}(4)$,
$\Psi_2=Z_3$ and $\Psi_3=Z_2$
Then from \eqref{Omegai6}, \eqref{Fi6} and \eqref{proj6},
$\Omega_1=\mathit{GF}(4)$, $\Omega_2=\{0, \omega, 2\omega\}$, $\Omega_3=\{0,
\omega^2\}$,
$F_1=\mathit{GF}(4)$, $F_2=\{k\omega+F_1, k\in Z_3\}$, $F_3=\{k\omega^2+F_2,
k\in Z_2\}$,
and for any $\gamma=\psi_0+\psi_1\omega+\psi_2\omega^2 \in F_3$,
$\rho_j(\gamma)=\psi_0+\cdots+\psi_{j-1}\omega^{j-1}$ for $j=1, 2, 3$,
where $\psi_b\in\Psi_{b+1}, b=0, 1, 2$.
Let
\begin{eqnarray*}
D_1&=&\left( \matrix{ 0& 0 & 0
\vspace*{2pt}\cr
0& 1 & x
\vspace*{2pt}\cr
0& x & x+1
\vspace*{2pt}\cr
0& x+1& 1 }
 \right), \qquad D_2=\left( %
\matrix{ 0 &0 & 0
\vspace*{2pt}\cr
0 &\omega& 2\omega
\vspace*{2pt}\cr
0 &2\omega& \omega}
 \right),\\
 D_3&=&\left(\matrix{ 0 & 0 & 0
\vspace*{2pt}\cr
0 & 0 & \omega^2
\vspace*{2pt}\cr
0 & \omega^2 & 0
\vspace*{2pt}\cr
0 & \omega^2 & \omega^2 }
 \right).
\end{eqnarray*}
Then
\begin{eqnarray*}
E_1 &=& D_1,\qquad  E_2=D_2
\oplus_cD_1, \qquad E_3=D_3
\oplus_c D_2\oplus_cD_1,
\\
\Delta^1_l&=&E_3\bigl(\bigl[4(l-1)+1\bigr]
\dvtx 4l\bigr),\qquad  l=1, \ldots, 12, \mbox{ and}
\\
\Delta^2_l&=&E_3\bigl(\bigl[12(l-1)+1\bigr]
\dvtx 12l\bigr),\qquad  l=1, \ldots, 4,
\end{eqnarray*}
which are listed in Table~\ref{exc2dm}.
%
\begin{table}
\caption{The array $E_3$ in Example \protect\ref{exc3}, where
{$E_1=E_3(1\dvtx 4)$, $E_2=E_3(1\dvtx 12)$,}
$\Delta^1_l=E_3([4(l-1)$ $+1]\dvtx 4l)$ for $l=1, \ldots, 12$,
and $\Delta^2_l=E_3([12(l-1)+1]\dvtx 12l)$ for $l=1, \ldots, 4$}
\label{exc2dm}
\begin{tabular*}{\textwidth}{@{\extracolsep{\fill}}lccccccc@{}}
\hline
\multicolumn{1}{@{}l}{\textbf{Row}}& \multicolumn{1}{c}{$\bolds{x_1}$} & \multicolumn{1}{c}{$\bolds{x_2}$} &
\multicolumn{1}{c}{$\bolds{x_3}$} & \multicolumn{1}{c}{\textbf{Row}} & \multicolumn{1}{c}{$\bolds{x_1}$} &
\multicolumn{1}{c}{$\bolds{x_2}$} & \multicolumn{1}{c@{}}{$\bolds{x_3}$} \\
\hline
\phantom{0}1 & 0 & 0 & 0 & 25 & 0 & $\omega^2$ & 0 \\
\phantom{0}2 & 0 & 1 & $x$ & 26 & 0 & $\omega^2$+1 & $x$ \\
\phantom{0}3 & 0 & $x$ & $x$+1 & 27 & 0 & $\omega^2$+$x$ & $x$+1 \\
\phantom{0}4 & 0 & $x$+1 & 1 & 28 & 0 & $\omega^2$+$x$+1 & 1 \\
\phantom{0}5 & 0 & $\omega$ & $2\omega$ & 29 & 0 & $\omega^2$+$\omega$ & $2\omega$
\\
\phantom{0}6 & 0 & $\omega$+1 & $2\omega$+$x$ & 30 & 0 & $\omega^2$+$\omega$+1 &
$2\omega$+$x$ \\
\phantom{0}7 & 0 & $\omega$+$x$ & $2\omega$+$x$+1 & 31 & 0 & $\omega^2$+$\omega
$+$x$ & $2\omega$+$x$+1 \\
\phantom{0}8 & 0 & $\omega$+$x$+1 & $2\omega$+1 & 32 & 0 & $\omega^2$+$\omega
$+$x$+1 & $2\omega$+1 \\
\phantom{0}9 & 0 & $2\omega$ & $\omega$ & 33 & 0 & $\omega^2$+$2\omega$ & $\omega$
\\
10 & 0 & $2\omega$+1 & $\omega$+$x$ & 34 & 0 & $\omega^2$+$2\omega$+1 &
$\omega$+$x$ \\
11 & 0 & $2\omega$+$x$ & $\omega$+$x$+1 & 35 & 0 & $\omega^2$+$2\omega
$+$x$ & $\omega$+$x$+1 \\
12 & 0 & $2\omega$+$x$+1 & $\omega$+1 & 36 & 0 & $\omega^2$+$2\omega
$+$x$+1 & $\omega$+1 \\
13 & 0 & 0 & $\omega^2$ & 37 & 0 & $\omega^2$ & $\omega^2$ \\
14 & 0 & 1 & $\omega^2$+$x$ & 38 & 0 & $\omega^2$+1 & $\omega^2$+$x$ \\
15 & 0 & $x$ & $\omega^2$+$x$+1 & 39 & 0 & $\omega^2$+$x$ & $\omega
^2$+$x$+1 \\
16 & 0 & $x$+1 & $\omega^2$+1 & 40 & 0 & $\omega^2$+$x$+1 & $\omega
^2$+1 \\
17 & 0 & $\omega$ & $\omega^2$+$2\omega$ & 41 & 0 & $\omega^2$+$\omega$
& $\omega^2$+$2\omega$ \\
18 & 0 & $\omega$+1 & $\omega^2$+$2\omega$+$x$ & 42 & 0 & $\omega
^2$+$\omega$+1 & $\omega^2$+$2\omega$+$x$ \\
19 & 0 & $\omega$+$x$ & $\omega^2$+$2\omega$+$x$+1& 43 & 0 & $\omega
^2$+$\omega$+$x$ & $\omega^2$+$2\omega$+$x$+1\\
20 & 0 & $\omega$+$x$+1 & $\omega^2$+$2\omega$+1 & 44 & 0 & $\omega
^2$+$\omega$+$x$+1 & $\omega^2$+$2\omega$+1 \\
21 & 0 & $2\omega$ & $\omega^2$+$\omega$ & 45 & 0 & $\omega^2$+$2\omega
$ & $\omega^2$+$\omega$ \\
22 & 0 & $2\omega$+1 & $\omega^2$+$\omega$+$x$ & 46 & 0 & $\omega
^2$+$2\omega$+1 & $\omega^2$+$\omega$+$x$ \\
23 & 0 & $2\omega$+$x$ & $\omega^2$+$\omega$+$x$+1& 47 & 0 & $\omega
^2$+$2\omega$+$x$ & $\omega^2$+$\omega$+$x$+1\\
24 & 0 & $2\omega$+$x$+1 & $\omega^2$+$\omega$+1 & 48 & 0 & $\omega
^2$+$2\omega$+$x$+1 & $\omega^2$+$\omega$+1 \\ \hline
\end{tabular*}                    \vspace*{-3pt}
\end{table}

It can be verified that:
\begin{longlist}[{(ii)}]
\item[{(i)}] $(E_1, E_2, E_3; \rho_1, \rho_2, \rho_3)$ is an NDM
with three layers,
where $\rho_j(E_i)$'s are difference matrices:
$\rho_1(E_1)=E_1$, $\rho_1(E_2)=(E_1', E_1', E_1')'$, $\rho_2(E_2)=E_2$,
$\rho_1(E_3)=(\underbrace{E_1', \ldots, E_1'} _{12})'$, $\rho
_2(E_3)=(E_2',E_2',E_2',E_2')'$
and $\rho_3(E_3)= E_3$;

\item[{(ii)}] $\rho_1(\Delta^1_l)=E_1$ for $l=1, \ldots, 12$,
$\rho_1(\Delta^2_l)=(E_1', E_1', E_1')'$ for $l=1, \ldots, 4$,
$\rho_2(\Delta^2_l)=E_2$ for $l=1, \ldots, 4$, which are all
difference matrices.\vspace*{-2pt}
\end{longlist}
\end{example}

\begin{remark} \label{rmk5}
Theorem~\ref{mthm3} provides a method for constructing NOAs and SOAs
from NDMs. The method can also be applied to generate NOAs and SOAs
using the NDMs obtained in Theorem~\ref{thmcnd} in a similar fashion
and the details are omitted.\vspace*{-2pt}
\end{remark}

\section{Generation of space-filling designs from NOAs and SOAs}\label{sec7}

We now discuss procedures for using the constructed NOAs and SOAs to
generate NSFDs and
SSFDs, respectively. Without loss of generality, we consider generating
space-filling designs from the NOAs and SOAs in Theorem~\ref{mthm1}.
Similar procedures can be carried out for other NOAs and SOAs.\vspace*{-2pt}

\subsection{Generation of NSFDs}

Qian, Tang and Wu (\citeyear{QTW}) proposed a meth\-od for generating NSFDs from
NOAs with two layers
and we extend their idea to generate NSFDs\vadjust{\goodbreak} with more than two layers.
We first introduce the definition of nested permutation with $I$ layers
[\citet{Q09}]. Let $Z_{s_{I}}=\{0, 1, \ldots, s_I-1\}$, we call $\pi_{\np
}=(\pi_{\np}(1), \ldots, \pi_{\np}(s_I))$ a \emph{nested permutation}
with $I$ layers on $Z_{s_{I}}$, if the $s_i$ elements of
$(\lfloor\pi_{\np}(1)s_i/s_I\rfloor, \ldots, \lfloor\pi_{\np
}(s_i)s_i/s_I \rfloor)$ is a
permutation on $Z_{s_i}=\{0, 1, \ldots, s_i-1\}$ for $i=1, \ldots, I$,
where $\lfloor z \rfloor$ denotes the largest integer no larger than
$z$ [\citet{Q09}]. Note that a necessary and sufficient condition for a
$\pi_{\np}$ to be a nest permutation is that precisely one of its first
$s_i$ entries falls within each of the $s_i$ sets defined by $\{0,
\ldots, s_I/s_i-1\}, \{s_I/s_i, \ldots, 2s_I/s_i-1\}, \ldots, \{(s_i-1)s_I/s_i,
\ldots, s_I-1\}$ for $i=1, \ldots, I$. \citet{Q09} presented an
algorithm for generating nested permutations with $I$ layers on $\{1,
2, \ldots, s_I\}$, which can be modified to generate nested
permutations with $I$ layers on $Z_{s_{I}}$, using the same uniform
permutations as in \citet{Q09}. Now we propose an algorithm using this
type of permutation to relabel the levels of $A_I$ and then obtain an NSFD.

\begin{algo}\label{algo2}
\textit{Step} 1. Take an NOA $(A_1, \ldots, A_I; \rho_1, \ldots, \rho_I)$ from
Theorem~\ref{mthm1} and let $\pi^l_{\np}$ be a nested permutation with
$I$ layers
on $Z_{s_{I}}$, $l=1, \ldots, (p^k-1)/(p-1)$.

\textit{Step} 2.
Relabel the levels of the $l$th column of $A_{I}$ according to
$\widetilde V (r)\longrightarrow\pi^l_{\np}(r)$ for $r=1, \ldots,
s_I$, and $l=1, \ldots, (p^k-1)/(p-1)$, where $ \widetilde
V=(\widetilde V(r))=V_{T_I}\oplus V_{T_{I-1}}\oplus \cdots\oplus
V_{T_1}$ [note that $\widetilde V$ is
different from the $V_{F_I}$ defined in \eqref{peiji1}]. Let $M_{I}$
be the resulting matrix.

\textit{Step} 3. Obtain an OA-based Latin hypercube $L_{I}$ from $M_{I}$.

\textit{Step} 4. Take $L_i$ to be the submatrix of $L_{I}$ consisting of
the first $s^k_i$ rows given by $L_i=L_{I}(1\dvtx s^k_i)$, for $i=1, \ldots, I-1$.
\end{algo}

\begin{thmm}\label{sfoa} The $(L_1, \ldots, L_I)$ is an NSFD with $I$
layers, where $L_i$ not only achieves stratification in any one
dimension, but also achieves stratification on
the $s_i\times s_i$ grids in any two dimensions for $i=1, \ldots, I$.
\end{thmm}

\begin{pf}
Note that $\rho_j(A_i)$ is an $\mathit{OA}(s^k_i, (p^k-1)/(p-1), s_j, 2)$ and
the entries of $F_i$ are relabeled with the first $s_i$ entries of $\pi
^l_{\np}$, where precisely one of these first $s_i$ entries falls
within each of the $s_i$ sets defined by $\{0, \ldots, s_I/s_i-1\}, \{
s_I/s_i, \ldots, 2s_I/s_i-1\}, \ldots, \{(s_i-1)s_I/s_i, \ldots, s_I-1\}
$, $1\leq j\leq i\leq I$ and $l=1, \ldots, (p^k-1)/(p-1)$. The
conclusions now follow.
\end{pf}

\begin{example}[(Example~\ref{me2} continued)]\label{esfoa}
Generate three nested permutations with three layers $\pi^1_{\np}=(4,
1, 2, 7, 6, 5, 3, 0)$, $\pi^2_{\np}=(5, 2, 0, 7, 3, 4, 1, 6)$, and $\pi
^3_{\np}=(2, 6, 1, 4, 3, 5, 7, 0)$ on $Z_8=\{0, \ldots, 7\}$. Note that
precisely one of the first $2^i$ entries of $\pi^l_{\np}$ falls within
each of the $2^i$ sets defined by $\{0,\ldots,2^{3-i}-1\}, \{
2^{3-i},\ldots,2\times2^{3-i}-1\}, \ldots, \{(2^i-1)2^{3-i}, \ldots,
2^{3}-1\}$, $i, l=1, 2, 3$. Relabel the levels of the $l$th column of
$A_3$ according to $\widetilde V(r)\longrightarrow \pi^l_{\np}(r),
r=1, \ldots, 8, l=1, 2, 3$, where $\widetilde V=(0, 1, x, x+1, x^2,
x^2+1, x^2+x, x^2+x+1)'$. The resulting matrix $M_{3}$ is given in
Table~\ref{mtb2}.\vadjust{\goodbreak} Use $M_3$ to obtain an OA-based Latin hypercube
$L_{3}$ listed in Table~\ref{emtb3}, and take $L_1$ and $L_2$ to be the
first four and sixteen rows of $L_3$, respectively. The bivariate
projections among $x_1, x_2, x_3$ of $L_3$ are plotted in Figure~\ref{mf1}, where the symbols
``$*$'', ``$+$'' and ``$\lozenge$'' denote the points
in $L_1$, $L_2\setminus L_1$ and $L_3\setminus L_2$, respectively.
The figure indicates that $L_i$ achieves stratification on the
$2^i\times2^i$ grids in any two dimensions for $i=1, 2, 3$.
%
\begin{table}
\caption{$M_3$ in Example \protect\ref{esfoa}}
\label{mtb2}
\begin{tabular*}{\textwidth}{@{\extracolsep{\fill}}lccccccc@{}}
\hline
\multicolumn{1}{@{}l}{\textbf{Row}} & \multicolumn{1}{c}{$\bolds{x_1}$} & \multicolumn{1}{c}{$\bolds{x_2}$} &
\multicolumn{1}{c}{$\bolds{x_3}$} &\multicolumn{1}{c}{\textbf{Row}} & \multicolumn{1}{c}{$\bolds{x_1}$} & \multicolumn{1}{c}{$\bolds{x_2}$} &
\multicolumn{1}{c@{}}{$\bolds{x_3}$}\\
\hline
\phantom{0}1 & 4 & 5 & 2 & 33 & 6 & 5 & 3 \\
\phantom{0}2 & 4 & 2 & 6 & 34 & 6 & 2 & 5 \\
\phantom{0}3 & 1 & 5 & 6 & 35 & 5 & 5 & 5 \\
\phantom{0}4 & 1 & 2 & 2 & 36 & 5 & 2 & 3 \\
\phantom{0}5 & 4 & 0 & 1 & 37 & 6 & 0 & 7 \\
\phantom{0}6 & 4 & 7 & 4 & 38 & 6 & 7 & 0 \\
\phantom{0}7 & 1 & 0 & 4 & 39 & 5 & 0 & 0 \\
\phantom{0}8 & 1 & 7 & 1 & 40 & 5 & 7 & 7 \\
\phantom{0}9 & 2 & 5 & 1 & 41 & 3 & 5 & 7 \\
10 & 2 & 2 & 4 & 42 & 3 & 2 & 0 \\
11 & 7 & 5 & 4 & 43 & 0 & 5 & 0 \\
12 & 7 & 2 & 1 & 44 & 0 & 2 & 7 \\
13 & 2 & 0 & 2 & 45 & 3 & 0 & 3 \\
14 & 2 & 7 & 6 & 46 & 3 & 7 & 5 \\
15 & 7 & 0 & 6 & 47 & 0 & 0 & 5 \\
16 & 7 & 7 & 2 & 48 & 0 & 7 & 3 \\
17 & 4 & 3 & 3 & 49 & 6 & 3 & 2 \\
18 & 4 & 4 & 5 & 50 & 6 & 4 & 6 \\
19 & 1 & 3 & 5 & 51 & 5 & 3 & 6 \\
20 & 1 & 4 & 3 & 52 & 5 & 4 & 2 \\
21 & 4 & 1 & 7 & 53 & 6 & 1 & 1 \\
22 & 4 & 6 & 0 & 54 & 6 & 6 & 4 \\
23 & 1 & 1 & 0 & 55 & 5 & 1 & 4 \\
24 & 1 & 6 & 7 & 56 & 5 & 6 & 1 \\
25 & 2 & 3 & 7 & 57 & 3 & 3 & 1 \\
26 & 2 & 4 & 0 & 58 & 3 & 4 & 4 \\
27 & 7 & 3 & 0 & 59 & 0 & 3 & 4 \\
28 & 7 & 4 & 7 & 60 & 0 & 4 & 1 \\
29 & 2 & 1 & 3 & 61 & 3 & 1 & 2 \\
30 & 2 & 6 & 5 & 62 & 3 & 6 & 6 \\
31 & 7 & 1 & 5 & 63 & 0 & 1 & 6 \\
32 & 7 & 6 & 3 & 64 & 0 & 6 & 2 \\
\hline
\end{tabular*}
\end{table}
%

\begin{table}
\caption{$L_3$ in Example \protect\ref{esfoa}, where $L_1=L_3(1\dvtx 4)$,
$L_2=L_3(1\dvtx 16)$}
\label{emtb3}
\begin{tabular*}{\textwidth}{@{\extracolsep{\fill}}lccccccc@{}}
\hline
\multicolumn{1}{@{}l}{\textbf{Row}} & \multicolumn{1}{c}{$\bolds{x_1}$} & \multicolumn{1}{c}{$\bolds{x_2}$} &
\multicolumn{1}{c}{$\bolds{x_3}$} &\multicolumn{1}{c}{\textbf{Row}} & \multicolumn{1}{c}{$\bolds{x_1}$} & \multicolumn{1}{c}{$\bolds{x_2}$} &
\multicolumn{1}{c@{}}{\textup{$\bolds{x_3}$}}\\
\hline
\phantom{0}1 & 39 & 44 & 17 & 33 & 51 & 47 & 28 \\
\phantom{0}2 & 38 & 19 & 49 & 34 & 52 & 17 & 41 \\
\phantom{0}3 & 12 & 40 & 50 & 35 & 44 & 46 & 42 \\
\phantom{0}4 & 13 & 18 & 21 & 36 & 46 & 21 & 24 \\
\phantom{0}5 & 34 & \phantom{0}7 & \phantom{0}8 & 37 & 54 & \phantom{0}2 & 62 \\
\phantom{0}6 & 33 & 58 & 33 & 38 & 49 & 62 & \phantom{0}7 \\
\phantom{0}7 & 10 & \phantom{0}5 & 39 & 39 & 43 & \phantom{0}0 & \phantom{0}0 \\
\phantom{0}8 & 11 & 61 & 14 & 40 & 40 & 59 & 58 \\
\phantom{0}9 & 22 & 41 & \phantom{0}9 & 41 & 24 & 42 & 60 \\
10 & 19 & 20 & 34 & 42 & 25 & 22 & \phantom{0}2 \\
11 & 60 & 45 & 36 & 43 & \phantom{0}2 & 43 & \phantom{0}1 \\
12 & 59 & 16 & 15 & 44 & \phantom{0}6 & 23 & 61 \\
13 & 23 & \phantom{0}4 & 23 & 45 & 30 & \phantom{0}1 & 30 \\
14 & 17 & 60 & 51 & 46 & 29 & 56 & 43 \\
15 & 61 & \phantom{0}3 & 52 & 47 & \phantom{0}5 & \phantom{0}6 & 40 \\
16 & 58 & 57 & 20 & 48 & \phantom{0}4 & 63 & 26 \\
17 & 35 & 28 & 27 & 49 & 53 & 31 & 16 \\
18 & 36 & 32 & 45 & 50 & 50 & 38 & 53 \\
19 & \phantom{0}8 & 25 & 46 & 51 & 47 & 27 & 48 \\
20 & 14 & 35 & 29 & 52 & 41 & 36 & 22 \\
21 & 32 & \phantom{0}9 & 63 & 53 & 55 & 13 & 10 \\
22 & 37 & 52 & \phantom{0}3 & 54 & 48 & 48 & 35 \\
23 & \phantom{0}9 & 15 & 6 & 55 & 45 & 11 & 37 \\
24 & 15 & 51 & 59 & 56 & 42 & 50 & 13 \\
25 & 16 & 30 & 56 & 57 & 27 & 24 & 12 \\
26 & 20 & 37 & \phantom{0}4 & 58 & 26 & 39 & 38 \\
27 & 62 & 26 & \phantom{0}5 & 59 & \phantom{0}3 & 29 & 32 \\
28 & 57 & 33 & 57 & 60 & \phantom{0}7 & 34 & 11 \\
29 & 18 & \phantom{0}8 & 31 & 61 & 28 & 10 & 19 \\
30 & 21 & 49 & 47 & 62 & 31 & 55 & 55 \\
31 & 56 & 14 & 44 & 63 & \phantom{0}1 & 12 & 54 \\
32 & 63 & 53 & 25 & 64 & \phantom{0}0 & 54 & 18 \\
\hline
\end{tabular*}
\end{table}
%

\begin{figure}

\includegraphics{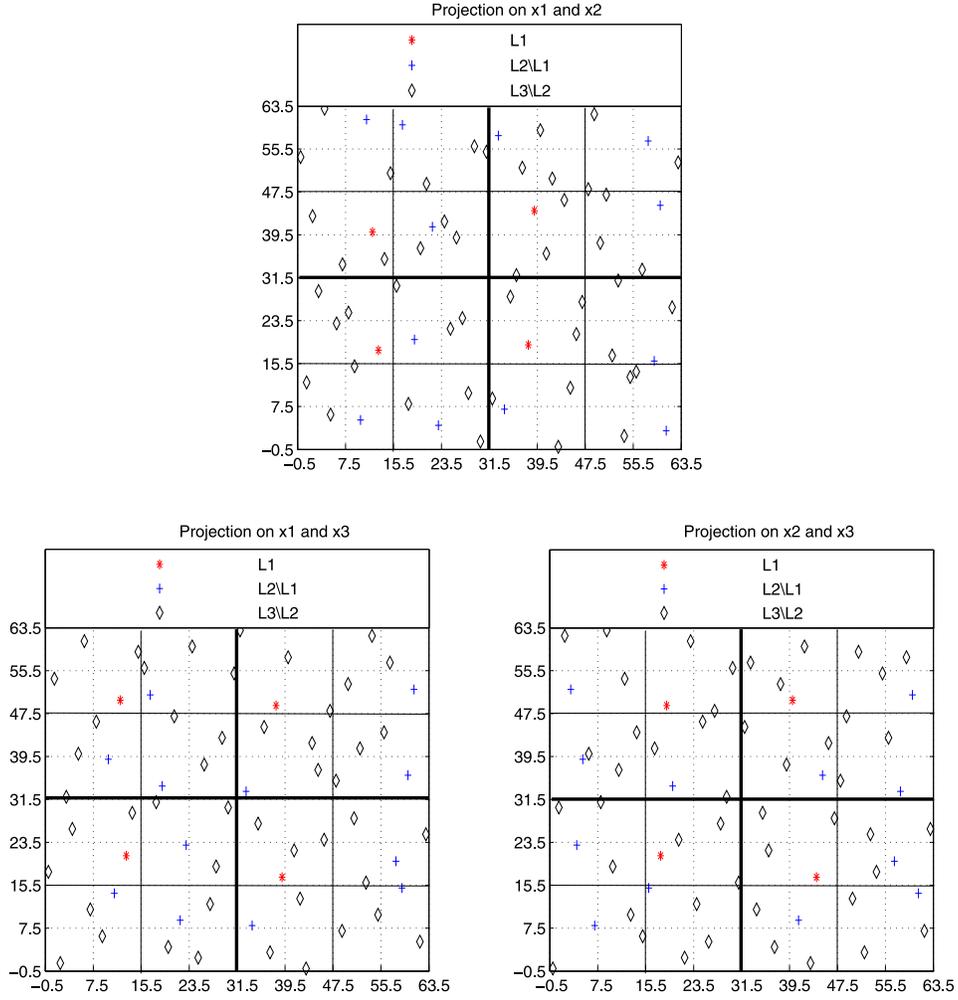}

\caption{Bivariate projections among $x_1, x_2, x_3$ of $L_3$ in Example
\protect\ref{esfoa}.} \label{mf1}\vspace*{-3pt}
\end{figure}
\end{example}

\subsection{Generation of SSFDs}

Qian and Wu (\citeyear{QW09}) proposed a method to obtain
SSFDs from SOAs. Here we
present a more flexible procedure that can use the SOAs constructed in
Sections~\ref{sec4}--\ref{sec6}
to generate a new class of SSFDs. Without loss of generality, consider
the SOAs constructed in Theorem~\ref{mthm1}.

\begin{algo}\label{algo3}
\textit{Step} 1.
Choose the values of $i,j,I$, where $1\leq j\leq i\leq I$. Suppose
$A_I$ and $(\Gamma^i_1, \ldots, \Gamma^i_{(s_I/s_i)^k}; \rho_j)$ are
constructed in Theorem~\ref{mthm1}. Relabel the $s_I$ levels of $A_I$
as $0, \ldots, s_I-1$
according to the following two stages:
\begin{longlist}[(ii)]
\item[(i)] Use the projection $\rho_j$ defined in \eqref{eq3} to
divide the $s_I$ levels into $s_j$ groups
\[
\Phi^j_{\alpha}=\bigl\{\gamma|\rho_j(\gamma)=
\alpha, \gamma\in F_I\bigr\}\qquad \mbox{for } \alpha\in F_j,
\]
each of size $q = s_I/s_j$.
\item[(ii)] Arbitrarily label the $s_j$ groups as groups $1, \ldots, s_j$,
and label the $q$ levels within the $g$th group as $(g-1)q, (g-1)q+1,
\ldots, gq-1$, for $g=1, \ldots, s_j$. This relabeling scheme can be
denoted by
%
\begin{equation}
\label{scheme} \bigl\{\Phi^j_{\alpha}|\alpha\in
F_j\bigr\} \longrightarrow \Lambda^j=\bigl\{
\Lambda^j_g| g=1, \ldots, s_j\bigr\},
\end{equation}
where $\Lambda^j_g=\{(g-1)q, (g-1)q+1, \ldots, gq-1 | q = s_I/s_j\}$.
\end{longlist}

\textit{Step} 2. Let $M$ be the design obtained by relabeling the levels
of $A_I$, and use $M$ to obtain an OA-based Latin hypercube $S$.

\textit{Step} 3. Partition $S$ into $(s_I/s_i)^k$ subarrays corresponding
to $\Gamma^i_1, \ldots,\break  \Gamma^i_{(s_I/s_i)^k}$, that is, $S=(S_1', \ldots,
S_{(s_I/s_i)^k}')'$ with $S_l=S([(l-1)s_i^k+1]\dvtx ls_i^k)$, $l=1, \ldots,
(s_I/s_i)^k$.
\end{algo}

\begin{thmm}\label{ssfoa}
For $S=(S_1', \ldots, S_{(s_I/s_i)^k}')'$ constructed in Algorithm \ref
{algo3}, $S$ achieves stratification on the $s_I\times s_I$ grids in
any two dimensions, and $S_l$ achieves stratification on the $s_j\times
s_j$ grids in any two dimensions\vadjust{\goodbreak} for $l=1, \ldots, (s_I/s_i)^k$. Thus,
$S=(S_1', \ldots, S_{(s_I/s_i)^k}')'$ is an SSFD with
$(s_I/s_i)^k$ slices.
\end{thmm}

\begin{pf}
By noting that $A_I$ and $\rho_j(\Gamma^i_l)$ for $l=1, \ldots,
(s_I/s_i)^k$ are all orthogonal arrays of strength two, and following
the relabeling scheme given above,
the conclusions hold.
\end{pf}

\begin{example}[(Example~\ref{me2} continued)]\label{essfoa}
(i) For $i=j=1$, we have $q=4, \Phi^1_{0}=\{\gamma| \rho_1(\gamma)=0,
\gamma\in F_3\}=\{0, x^2, x, x^2+x\}, \Phi^1_{1}=\{1, x^2+1, x+1,
x^2+x+1\}, \Lambda^1_1=\{0, 1, 2, 3\}$ and $\Lambda^1_2=\{4, 5, 6, 7\}
$. Arbitrarily relabel the levels of $A_3$ in Table~\ref{mtb1}
according to the scheme given in Step 1 as follows:
\[
\bigl\{\bigl\{0,x^2,x,x^2+x\bigr\},\bigl
\{1,x^2+1,x+1,x^2+x+1\bigr\}\bigr\}\longrightarrow \bigl\{
\{0,1,2,3\},\{4,5,6,7\}\bigr\},
\]
and then obtain an OA-based Latin hypercube $S$. Let $S_l=S([4(l-1)+1]
\dvtx 4l), l=1, \ldots, 16$. Note that $S$ achieves stratification on the
$8\times8$ grids in any two dimensions, $S_l$ achieves stratification
on the $2\times2$ grids in any two dimensions, and $S=(S_1', \ldots,
S_{16}')'$ is an SSFD with 16 slices.

(ii) For $i=j=2$, we have $q=2,
\Phi^2_{0}=\{0, x^2\},
\Phi^2_{x}=\{x, x^2+x\},
\Phi^2_{1}=\{1, x^2+1\},
\Phi^2_{x+1}=\{x+1, x^2+x+1\},
\Lambda^2_1=\{0, 1\},
\Lambda^2_2=\{2, 3\},
\Lambda^2_3=\{4, 5\}$ and $\Lambda^2_4=\{6, 7\}$.
Relabel the levels of $A_3$ according to
\begin{eqnarray*}
&&\bigl\{\bigl\{0,x^2\bigr\},\bigl\{x, x^2+x\bigr\},\bigl
\{1,x^2+1\bigr\},\bigl\{x+1,x^2+x+1\bigr\}\bigr\}
\\
&&\qquad\longrightarrow\bigl\{\{0,1\},\{2,3\},\{4,5\},\{6,7\}\bigr\},
\end{eqnarray*}
to obtain an OA-based Latin hypercube $S=(S_1', \ldots, S_{4}')'$, where
$S_l=\break S([16(l-1)+1]\dvtx 16l), l=1, \ldots, 4$. Similarly, $S$ achieves
stratification on the $8\times8$ grids in any two dimensions, $S_l$
achieves stratification on the $4\times4$ grids in any two dimensions,
and $S=(S_1', \ldots, S_{4}')'$ is an SSFD with 4 slices.
\end{example}

\begin{remark}\label{rmk6}
If we relabel the levels of $A_3$ according to
%
\begin{eqnarray}
\label{scheme2}
\bigl\{\bigl\{0, x^2\bigr
\}, \bigl\{x, x^2+x\bigr\}\bigr\} &\longrightarrow&\bigl\{\{0, 1\}, \{2,
3\}\bigr\}\quad\mbox{and}
\nonumber
\\[-8pt]
\\[-8pt]
\nonumber
\bigl\{\bigl\{1, x^2+1\bigr\}, \bigl\{x+1, x^2+x+1\bigr
\}\bigr\}& \longrightarrow&\bigl\{\{4, 5\}, \{6, 7\}\bigr\},
\end{eqnarray}
in Example~\ref{essfoa}, then by Theorem~\ref{ssfoa}, we have:
\begin{longlist}[(a)]
\item[(a)] $S$ can be partitioned into 16 slices, $S([4(l-1)+1]\dvtx 4l)$
for $l=1, \ldots, 16$, each of which achieves stratification on
the $2\times2$ grids in any two dimensions;

\item[(b)] $S$ can be partitioned into 4 slices, $S([16(l-1)+1]\dvtx 16l)$
for $l=1, \ldots, 4$, each of which achieves stratification on
the $4\times4$ grids in any two dimensions;\vadjust{\goodbreak}

\item[(c)] $S$ achieves stratification on the $8\times8$ grids in any
two dimensions;

\item[(d)] $S$ is an SSFD that can be sliced into 4 or 16 slices.
\end{longlist}
Therefore, under the same relabel scheme \eqref{scheme2}, $S$ can be
used to conduct computer experiments with qualitative factors of 4 and
16 distinct level combinations, respectively. A further discussion on
$S$ will be found in Example~\ref{essfoa2}.\vadjust{\goodbreak}
\end{remark}

Inspired by Remark~\ref{rmk6}, we now propose a new construction of
SSFDs from SOAs which can generate SSFDs with different numbers of
slices simultaneously. A new permutation is needed. We call
$\pi_{\slp}=(\pi_{\slp}(1), \ldots, \pi_{\slp}(s_I))$ a \emph{sliced
permutation}
with $I$ layers on $Z_{s_{I}}$, if
\[
\bigl\{\pi_{\slp}\bigl((g-1)q+1\bigr), \pi_{\slp}
\bigl((g-1)q+2\bigr), \ldots, \pi_{\slp
}(gq) \bigr\}\in
\Lambda^j
\]
for $j=1, \ldots, I-1$, $g=1, \ldots, s_j$ and $q=s_I/s_j$, where $
\Lambda^j$ is
defined in \eqref{scheme}.

\begin{algo}\label{algo4}
\textit{Step} 1.
Suppose $A_I$ is constructed in Theorem~\ref{mthm1} and $\pi^l_{\slp}$
is a sliced permutation with $I$ layers on $Z_{s_{I}}$, $l=1, \ldots,
(p^k-1)/(p-1)$.

\textit{Step} 2.
Relabel the levels of the $l$th column of $A_{I}$ according to
$V_{F_I}(r)\longrightarrow\pi^l_{\slp}(r)$ for $r=1, \ldots, s_I$, and
$l=1, \ldots, (p^k-1)/(p-1)$, where $V_{F_I}=V_{T_1}\oplus
V_{T_{2}}\oplus\cdots\oplus V_{T_I}$
defined in \eqref{peiji1}. Let $M$ be the resulting matrix.

\textit{Step} 3. Obtain an OA-based Latin hypercube $S$ from $M$.

\textit{Step} 4. For $i=1, \ldots, I-1$, partition $S$ into $(s_I/s_i)^k$
subarrays with\vspace*{1pt}
an equal number of rows, that is, $S=((S_1^{i})', \ldots,
(S_{(s_I/s_i)^k}^{i})')'$ with $S_l^{i}=S([(l-1)s_i^k+1]\dvtx ls_i^k)$ for
$l=1, \ldots, (s_I/s_i)^k$.
\end{algo}

\begin{thmm}\label{ssfoa2}
For $S=((S_1^{i})', \ldots, (S_{(s_I/s_i)^k}^{i})')'$ constructed in
Algorithm \ref{algo4}, $S_l^{i}$ achieves stratification on the
$s_j\times s_j$ grids in any two dimensions, for $l=1, \ldots,
(s_I/s_i)^k$ and $1\leq j\leq i \leq I$. Thus, $S=((S_1^{i})', \ldots,
(S_{(s_I/s_i)^k}^{i})')'$ is an SSFD with $(s_I/s_i)^k$ slices, for
$i=1, \ldots, I-1$.
\end{thmm}

\begin{pf}
For any $\alpha\in F_I$, let ${\alpha}_l$ denote the corresponding element
in $\pi^l_{\slp}$ under the relabeling $V_{F_I} \longrightarrow\pi
^l_{\slp},
l=1, \ldots, (p^k-1)/(p-1)$. Since $A_I$ and $\rho_j(\Gamma^i_l)$ for
$l=1, \ldots, (s_I/s_i)^k$ and $1\leq j\leq i\leq I-1$ are all
orthogonal arrays of strength two, it suffices to prove that for any
$\alpha, \beta\in F_j$ with $\alpha\neq\beta$,
${\alpha}_l$ and ${\beta}_l$ fall in different sets defined by $\{0, 1,
\ldots, q-1\}, \{q, q+1, \ldots, 2q-1\}, \ldots, \{(s_j-1)q,
(s_j-1)q+1, \ldots, s_jq-1\}$, where $q=s_I/s_j$.
Note that $V_{F_I}=V_{T_1}\oplus V_{T_{2}}\oplus\cdots\oplus
V_{T_I}=V_{F_j}\oplus(V_{T_{j+1}} \oplus\cdots\oplus V_{T_I})$ and
the first element of $V_{T_i}$ is 0, $i=1, \ldots, I$, then $\alpha,
\beta\in \{V_{F_I}(g) | g=1, q+1, 2q+1, \ldots, (s_j-1)q+1 \}
$. Suppose $\alpha=V_{F_I}(c_1q+1), \beta=V_{F_I}(c_2q+1), c_1, c_2=0,
1, \ldots, s_j-1$, and $c_1\neq c_2$. Then ${\alpha}_l=\pi^l_{\slp
}(c_1q+1)$ and ${\beta}_l=\pi^l_{\slp}(c_2q+1)$ and, therefore, ${\alpha
}_l\in\Lambda^j_{d_1}, {\beta}_l\in\Lambda^j_{d_2}$ for some $d_1,
d_2=1, 2, \ldots, s_j$ and $d_1\neq d_2$
(this is because $|(c_1q+1)-(c_2q+1)|=|(c_1-c_2)q|\geq q$), and ${\alpha
}_l$ and ${\beta}_l$ fall in different sets defined by $\{0, 1, \ldots,
q-1\}, \{q, q+1, \ldots, 2q-1\}, \ldots, \{(s_j-1)q, (s_j-1)q+1, \ldots,
s_jq-1\}$.\vadjust{\goodbreak}~%
\end{pf}

\begin{table}
\def\arraystretch{0.99}
\caption{The array $M$ in Example \protect\ref{essfoa2}}
\label{emtb33}
\begin{tabular*}{\textwidth}{@{\extracolsep{\fill}}lccccccc@{}}
\hline
\multicolumn{1}{@{}l}{\textbf{Row}}& \multicolumn{1}{c}{$\bolds{x_1}$} & \multicolumn{1}{c}{$\bolds{x_2}$} &
\multicolumn{1}{c}{$\bolds{x_3}$} & \multicolumn{1}{c}{\textbf{Row}} & \multicolumn{1}{c}{$\bolds{x_1}$} & \multicolumn{1}{c}{$\bolds{x_2}$} & \multicolumn{1}{c@{}}{$\bolds{x_3}$} \\
\hline
\phantom{0}1 & 0 & 7 & 0 & 33 & 1 & 7 & 1 \\
\phantom{0}2 & 0 & 1 & 4 & 34 & 1 & 1 & 5 \\
\phantom{0}3 & 7 & 7 & 4 & 35 & 6 & 7 & 5 \\
\phantom{0}4 & 7 & 1 & 0 & 36 & 6 & 1 & 1 \\
\phantom{0}5 & 0 & 5 & 3 & 37 & 1 & 5 & 2 \\
\phantom{0}6 & 0 & 2 & 7 & 38 & 1 & 2 & 6 \\
\phantom{0}7 & 7 & 5 & 7 & 39 & 6 & 5 & 6 \\
\phantom{0}8 & 7 & 2 & 3 & 40 & 6 & 2 & 2 \\
\phantom{0}9 & 2 & 7 & 3 & 41 & 3 & 7 & 2 \\
10 & 2 & 1 & 7 & 42 & 3 & 1 & 6 \\
11 & 5 & 7 & 7 & 43 & 4 & 7 & 6 \\
12 & 5 & 1 & 3 & 44 & 4 & 1 & 2 \\
13 & 2 & 5 & 0 & 45 & 3 & 5 & 1 \\
14 & 2 & 2 & 4 & 46 & 3 & 2 & 5 \\
15 & 5 & 5 & 4 & 47 & 4 & 5 & 5 \\
16 & 5 & 2 & 0 & 48 & 4 & 2 & 1 \\
17 & 0 & 6 & 1 & 49 & 1 & 6 & 0 \\
18 & 0 & 0 & 5 & 50 & 1 & 0 & 4 \\
19 & 7 & 6 & 5 & 51 & 6 & 6 & 4 \\
20 & 7 & 0 & 1 & 52 & 6 & 0 & 0 \\
21 & 0 & 4 & 2 & 53 & 1 & 4 & 3 \\
22 & 0 & 3 & 6 & 54 & 1 & 3 & 7 \\
23 & 7 & 4 & 6 & 55 & 6 & 4 & 7 \\
24 & 7 & 3 & 2 & 56 & 6 & 3 & 3 \\
25 & 2 & 6 & 2 & 57 & 3 & 6 & 3 \\
26 & 2 & 0 & 6 & 58 & 3 & 0 & 7 \\
27 & 5 & 6 & 6 & 59 & 4 & 6 & 7 \\
28 & 5 & 0 & 2 & 60 & 4 & 0 & 3 \\
29 & 2 & 4 & 1 & 61 & 3 & 4 & 0 \\
30 & 2 & 3 & 5 & 62 & 3 & 3 & 4 \\
31 & 5 & 4 & 5 & 63 & 4 & 4 & 4 \\
32 & 5 & 3 & 1 & 64 & 4 & 3 & 0 \\
\hline
\end{tabular*}
\end{table}
%

\begin{figure}

\includegraphics{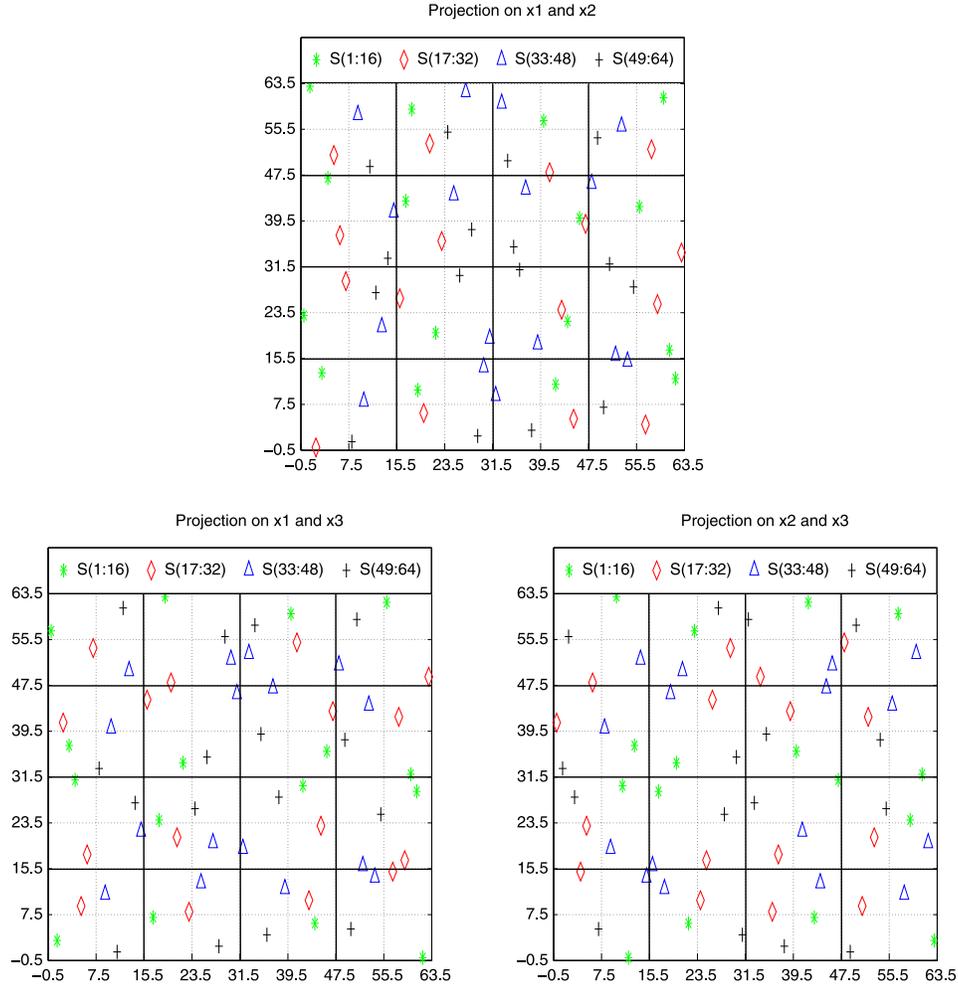}

\caption{Bivariate projections among $x_1$, $x_2$ and $x_3$ of $S$ in Example
\protect\ref{essfoa2}.}\label{mf2}
\end{figure}

\begin{example}[(Example~\ref{me2} continued)]\label{essfoa2}
Generate three sliced permutations with three layers $\pi^1_{\slp}=(0,
1, 2, 3, 7, 6, 5, 4), \pi^2_{\slp}=(7, 6, 5, 4, 1, 0, 2, 3)$ and $\pi
^3_{\slp}=(0, 1, 3, 2, 4, 5, 7, 6)$ on $Z_8$. Note that
\[
\bigl\{\pi^l_{\slp}\bigl(r2^{3-j}+1\bigr), \ldots,
\pi^l_{\slp
}\bigl((r+1)2^{3-j}\bigr) \bigr\}\in
\Lambda^j
\]
for $r=0, 1, \ldots, 2^{j}-1$ and $j=1, 2$, where $\Lambda^1= \{\{
0, 1, 2, 3\}, \{4, 5, 6, 7\} \}$ and $\Lambda^2= \{\{0, 1\}, \{
2, 3\}, \{4, 5\}, \{6, 7\} \}$. Relabel the levels of the $l$th
column of $A_3$ according to $V_{F_I}(r)\longrightarrow\pi^l_{\slp
}(r), r=1, \ldots, 8, l=1, 2, 3$, where $V_{F_I}=\{0, x^2, x,\break  x+x^2, 1,
x^2+1, x+1, x+x^2+1\}$. Denote the resulting matrix by $M$ in Table~\ref{emtb33}, and use $M$ to obtain an OA-based Latin hypercube $S$ given
in columns $x_1, x_2$ and $x_3$ in Table~\ref{mtb3}. Note that
$S([4^i(l-1)+1]\dvtx 4^i l)$ achieves stratification on the $2^i\times2^i$
grids in any
two dimensions for $l=1, 2, \ldots, 4^{3-i}$ and $i=1, 2$; see Figure~\ref{mf2} for an illustration, where for brevity, we only plot the
bivariate projections of
$S([16(l-1)+1]\dvtx 16l)$ for $l=1, \ldots, 4$.
%

\begin{table}
\caption{$\mathit{OA}(16, 2^34^3, 2)$}
\label{mtb20}
\begin{tabular*}{\textwidth}{@{\extracolsep{\fill}}lccccc@{}}
\hline
\textbf{1} & \textbf{2} & \textbf{3} & \textbf{4} & \textbf{5} & \textbf{6} \\
\hline
0 & 0 & 0 & 0 & 0 & 0 \\
0 & 0 & 0 & 0 & 3 & 3 \\
0 & 0 & 0 & 3 & 1 & 2 \\
0 & 0 & 0 & 3 & 2 & 1 \\
0 & 1 & 1 & 2 & 0 & 2 \\
0 & 1 & 1 & 2 & 3 & 1 \\
0 & 1 & 1 & 1 & 2 & 3 \\
0 & 1 & 1 & 1 & 1 & 0 \\
1 & 0 & 1 & 2 & 2 & 0 \\
1 & 0 & 1 & 2 & 1 & 3 \\
1 & 0 & 1 & 1 & 0 & 1 \\
1 & 0 & 1 & 1 & 3 & 2 \\
1 & 1 & 0 & 0 & 2 & 2 \\
1 & 1 & 0 & 0 & 1 & 1 \\
1 & 1 & 0 & 3 & 0 & 3 \\
1 & 1 & 0 & 3 & 3 & 0 \\
\hline
\end{tabular*}
\end{table}

The design in Table~\ref{mtb3} consists of two parts: the SSFD $S$
(columns $x_1, x_2, x_3$) obtained in Example~\ref{essfoa2} for
arranging quantitative factors, and
an $\mathit{OA}(16, 2^34^3)$ with replicate runs (the last six columns) for
arranging qualitative factors, where the original $\mathit{OA}(16, 2^34^3)$ is
listed in Table~\ref{mtb20}.
Note that $S$ possesses properties:
(i) if $S$ is partitioned into 4 slices with 16 runs in each slice,
then each slice achieves stratification on the $4\times4$ grids in any
two dimensions;
(ii) if $S$ is partitioned into 16 slices with 4 runs in each slice,
then each slice achieves stratification on the $2\times2$ grids in any
two dimensions.
Therefore, for the design in Table~\ref{mtb3},
(i) for any level combination of the three two-level qualitative
factors, the design points for the quantitative factors achieve
stratification on the $4\times4$ grids in any two dimensions;
(ii) for any level combination of the three four-level qualitative
factors, the design points for the quantitative factors achieve
stratification on the $2\times2$ grids in any two dimensions;
(iii) it possesses good space-filling properties when collapsed over
the qualitative factors. Hence, the design in Table~\ref{mtb3} is
suitable for conducting a computer
experiment with three quantitative factors and six qualitative factors,
where three of them have 2 levels and another three have 4 levels.
\end{example}

\begin{table}
\caption{Design with qualitative and quantitative factors,
where columns $x_1,x_2,x_3$ are
quantitative ones, $x_4,x_5,x_6$ are 2-level qualitative
ones, and $x_7,x_8,x_9$ are 4-level qualitative ones}
\label{mtb3}
\begin{tabular*}{\textwidth}{@{\extracolsep{\fill}}lccccccccc@{}}
\hline
\multicolumn{1}{@{}l}{\textbf{Row}} & \multicolumn{1}{c}{$\bolds{x_1}$} & \multicolumn{1}{c}{$\bolds{x_2}$} &
\multicolumn{1}{c}{$\bolds{x_3}$} & \multicolumn{1}{c}{$\bolds{x_4}$} & \multicolumn{1}{c}{$\bolds{x_5}$} &
\multicolumn{1}{c}{$\bolds{x_6}$} & \multicolumn{1}{c}{$\bolds{x_7}$} & \multicolumn{1}{c}{$\bolds{x_8}$} &
\multicolumn{1}{c@{}}{$\bolds{x_9}$} \\
\hline
\phantom{0}1 & \phantom{0}1 & 63 & \phantom{0}3 & 0 & 0 & 0 & 0 & 0 & 0 \\
\phantom{0}2 & \phantom{0}3 & 13 & 37 & 0 & 0 & 0 & 0 & 0 & 0 \\
\phantom{0}3 & 60 & 61 & 32 & 0 & 0 & 0 & 0 & 0 & 0 \\
\phantom{0}4 & 62 & 12 & \phantom{0}0 & 0 & 0 & 0 & 0 & 0 & 0 \\
\phantom{0}5 & \phantom{0}4 & 47 & 31 & 0 & 0 & 0 & 0 & 3 & 3 \\
\phantom{0}6 & \phantom{0}0 & 23 & 57 & 0 & 0 & 0 & 0 & 3 & 3 \\
\phantom{0}7 & 56 & 42 & 62 & 0 & 0 & 0 & 0 & 3 & 3 \\
\phantom{0}8 & 61 & 17 & 29 & 0 & 0 & 0 & 0 & 3 & 3 \\
\phantom{0}9 & 18 & 59 & 24 & 0 & 0 & 0 & 3 & 1 & 2 \\
10 & 19 & 10 & 63 & 0 & 0 & 0 & 3 & 1 & 2 \\
11 & 40 & 57 & 60 & 0 & 0 & 0 & 3 & 1 & 2 \\
12 & 42 & 11 & 30 & 0 & 0 & 0 & 3 & 1 & 2 \\
13 & 17 & 43 & \phantom{0}7 & 0 & 0 & 0 & 3 & 2 & 1 \\
14 & 22 & 20 & 34 & 0 & 0 & 0 & 3 & 2 & 1 \\
15 & 46 & 40 & 36 & 0 & 0 & 0 & 3 & 2 & 1 \\
16 & 44 & 22 & \phantom{0}6 & 0 & 0 & 0 & 3 & 2 & 1 \\
17 & \phantom{0}5 & 51 & \phantom{0}9 & 0 & 1 & 1 & 2 & 0 & 2 \\
18 & \phantom{0}2 & \phantom{0}0 & 41 & 0 & 1 & 1 & 2 & 0 & 2 \\
19 & 58 & 52 & 42 & 0 & 1 & 1 & 2 & 0 & 2 \\
20 & 57 & \phantom{0}4 & 15 & 0 & 1 & 1 & 2 & 0 & 2 \\
21 & \phantom{0}6 & 37 & 18 & 0 & 1 & 1 & 2 & 3 & 1 \\
22 & \phantom{0}7 & 29 & 54 & 0 & 1 & 1 & 2 & 3 & 1 \\
23 & 63 & 34 & 49 & 0 & 1 & 1 & 2 & 3 & 1 \\
24 & 59 & 25 & 17 & 0 & 1 & 1 & 2 & 3 & 1 \\
25 & 21 & 53 & 21 & 0 & 1 & 1 & 1 & 2 & 3 \\
26 & 20 & \phantom{0}6 & 48 & 0 & 1 & 1 & 1 & 2 & 3 \\
27 & 41 & 48 & 55 & 0 & 1 & 1 & 1 & 2 & 3 \\
28 & 45 & \phantom{0}5 & 23 & 0 & 1 & 1 & 1 & 2 & 3 \\
29 & 23 & 36 & \phantom{0}8 & 0 & 1 & 1 & 1 & 1 & 0 \\
30 & 16 & 26 & 45 & 0 & 1 & 1 & 1 & 1 & 0 \\
31 & 47 & 39 & 43 & 0 & 1 & 1 & 1 & 1 & 0 \\
32 & 43 & 24 & 10 & 0 & 1 & 1 & 1 & 1 & 0 \\
33 & \phantom{0}9 & 58 & 11 & 1 & 0 & 1 & 2 & 2 & 0 \\
34 & 10 & \phantom{0}8 & 40 & 1 & 0 & 1 & 2 & 2 & 0 \\
35 & 53 & 56 & 44 & 1 & 0 & 1 & 2 & 2 & 0 \\
36 & 54 & 15 & 14 & 1 & 0 & 1 & 2 & 2 & 0 \\
37 & 15 & 41 & 22 & 1 & 0 & 1 & 2 & 1 & 3 \\
38 & 13 & 21 & 50 & 1 & 0 & 1 & 2 & 1 & 3 \\
39 & 48 & 46 & 51 & 1 & 0 & 1 & 2 & 1 & 3 \\
40 & 52 & 16 & 16 & 1 & 0 & 1 & 2 & 1 & 3 \\
41 & 27 & 62 & 20 & 1 & 0 & 1 & 1 & 0 & 1 \\
42 & 30 & 14 & 52 & 1 & 0 & 1 & 1 & 0 & 1 \\
43 & 33 & 60 & 53 & 1 & 0 & 1 & 1 & 0 & 1 \\
44 & 32 & \phantom{0}9 & 19 & 1 & 0 & 1 & 1 & 0 & 1 \\
\hline
\end{tabular*}
\end{table}
\setcounter{table}{7}
\begin{table}
\caption{(Continued)}
\begin{tabular*}{\textwidth}{@{\extracolsep{\fill}}lccccccccc@{}}
\hline
\multicolumn{1}{@{}l}{\textbf{Row}} & \multicolumn{1}{c}{$\bolds{x_1}$} & \multicolumn{1}{c}{$\bolds{x_2}$} &
\multicolumn{1}{c}{$\bolds{x_3}$} & \multicolumn{1}{c}{$\bolds{x_4}$} & \multicolumn{1}{c}{$\bolds{x_5}$} &
\multicolumn{1}{c}{$\bolds{x_6}$} & \multicolumn{1}{c}{$\bolds{x_7}$} & \multicolumn{1}{c}{$\bolds{x_8}$} &
\multicolumn{1}{c@{}}{$\bolds{x_9}$} \\
\hline
45 & 25 & 44 & 13 & 1 & 0 & 1 & 1 & 3 & 2 \\
46 & 31 & 19 & 46 & 1 & 0 & 1 & 1 & 3 & 2 \\
47 & 37 & 45 & 47 & 1 & 0 & 1 & 1 & 3 & 2 \\
48 & 39 & 18 & 12 & 1 & 0 & 1 & 1 & 3 & 2 \\
49 & 11 & 49 & 1 & 1 & 1 & 0 & 0 & 2 & 2 \\
50 & \phantom{0}8 & \phantom{0}1 & 33 & 1 & 1 & 0 & 0 & 2 & 2 \\
51 & 49 & 54 & 38 & 1 & 1 & 0 & 0 & 2 & 2 \\
52 & 50 & \phantom{0}7 & \phantom{0}5 & 1 & 1 & 0 & 0 & 2 & 2 \\
53 & 14 & 33 & 27 & 1 & 1 & 0 & 0 & 1 & 1 \\
54 & 12 & 27 & 61 & 1 & 1 & 0 & 0 & 1 & 1 \\
55 & 51 & 32 & 59 & 1 & 1 & 0 & 0 & 1 & 1 \\
56 & 55 & 28 & 25 & 1 & 1 & 0 & 0 & 1 & 1 \\
57 & 24 & 55 & 26 & 1 & 1 & 0 & 3 & 0 & 3 \\
58 & 29 & \phantom{0}2 & 56 & 1 & 1 & 0 & 3 & 0 & 3 \\
59 & 34 & 50 & 58 & 1 & 1 & 0 & 3 & 0 & 3 \\
60 & 38 & \phantom{0}3 & 28 & 1 & 1 & 0 & 3 & 0 & 3 \\
61 & 28 & 38 & \phantom{0}2 & 1 & 1 & 0 & 3 & 3 & 0 \\
62 & 26 & 30 & 35 & 1 & 1 & 0 & 3 & 3 & 0 \\
63 & 35 & 35 & 39 & 1 & 1 & 0 & 3 & 3 & 0 \\
64 & 36 & 31 & \phantom{0}4 & 1 & 1 & 0 & 3 & 3 & 0 \\
\hline
\end{tabular*}
\end{table}

We have provided some new constructions of NSFDs and SSFDs based on
NOAs and SOAs of strength two, respectively. Better NSFDs and SSFDs can
be obtained by using NOAs and SOAs with strength greater than two. See
Remarks \ref{rmk3} and \ref{rmk4}, Theorems \ref{bush} and \ref{thml}
and Corollary~\ref{coroc2}.

\section{Comparisons and concluding remarks}\label{sec8}

The families of NSFDs constructed by the existing methods are limited
to two layers, with the exception of \citet{HQ}. The
method of \citet{HQ} is based on the infinite $(t,
s)$-sequences which are more difficult to obtain than the orthogonal
arrays used in our methods. Here are some comparisons between our
methods and the existing constructions.

\citet{QTW} (QTW) and Qian, Ai and Wu (\citeyear{QAW}) (QAW)
presented several methods for constructing NSFDs with two layers from
NOAs and NDMs. NSFDs with more than two layers cannot be constructed by
using their methods. The technical reason is that the modulus
projection used in \citet{QTW} cannot be extended to
covering more than two layers, as argued in Section~\ref{sec32}. The
subgroup projection presented in this paper
is different and more general, and it has been used to generate more
NSFDs which can accommodate nesting with an arbitrary number of layers
and are more flexible in run size. \citet{QA} (QA) proposed some
construction methods for NOAs and NDMs with two layers based on Galois
fields and incomplete pairwise orthogonal Latin squares. \citet{Q09}
presented a method for constructing nested Latin hypercube designs, but
the resulting designs can achieve stratification only in one dimension.
Thus, we only present the comparisons among QTW, QAW, QA and our
proposed methods (SLQ). The comparison among QAW, QA and SLQ for the
construction of NDMs with two layers, and the comparison among QTW,
QAW, QA and SLQ for the construction of NOAs with two layers, are
listed in Tables~\ref{comparison1} and \ref{comparison2}, respectively.
Since the construction of incomplete pairwise orthogonal Latin squares
is still an open problem, thus we only tabulate
the results obtained based on Galois fields in QA. In addition, QAW
and the present paper presented several indirect methods to obtain NOAs
based on existing NOAs or NDMs, for example,
Theorems 4, 5 in QAW and Theorem~\ref{mthm3} in the present paper.
In Tables~\ref{comparison1} and \ref{comparison2}, we only tabulate the
NOAs and NDMs that can be directly constructed. Moreover, Tables~\ref{SNDMS} and \ref{SNOAS} tabulate some construction results of the
proposed methods for designs with more than two
layers.

%
\begin{table}
\tabcolsep=0pt
\caption{Comparisons among the $\mathit{NDM}(D_1, D; \rho_1, \rho_2)$'s
constructed by QAW, QA
and SLQ}
\label{comparison1}
\begin{tabular*}{\textwidth}{@{\extracolsep{\fill}}lcccc@{}}
\hline
\multicolumn{2}{@{}l}{\textbf{Methods}} & $\bolds{\rho_1(D_1)}$ & $\bolds{D}$ & \textbf{Constraints\tabnoteref{tt9}}\\
\hline
QAW&I &$D(p^m, p^2, p^m)$&$D(p^{m+1}, p^2, p^{m+1})$&$m\geq2$\\
&II&$D(p^m, p^2, p^m)$&$D(p^{m+2}, p^2, p^{m+2})$&$p=2, 3$\\
 & III &$D(p^{m+1}, p^3, p^m)$&$D(p^{m+2}, p^3, p^{m+2})$&\\
&IV&$D(p^{m+1}, p^3, p^m)$&$D(p^{m+3}, p^3, p^{m+3})$&\\
&V &$D(p^{m+2}, p^4, p^m)$&$D(p^{m+3}, p^4, p^{m+3})$&\\[3pt]
\multicolumn{2}{@{}l}{QA} &$D(p^{u_1}, p^{u_1}, p^{u_1})$ & $D(p^{u_2},
p^{u_1}, p^{u_2})$& $u_1<u_2, u_1|u_{2}$\\ [3pt]
SLQ & Theorem~\ref{mthm3}& $D(lp^{u_2}, p^{u_1}, p^{u_2})$ & $D(p^{u_3},
p^{u_1}, p^{u_3})$& $u_i|u_{i+1}, i=1, 2$,\\
&&&& $u_2<u_3$, $l<
p^{u_3-u_2}$\\
&Theorem~\ref{thmcnd} & $D(lr_1, c, p^{u_1})$ &$D(r_1r_2, c, p^{u_1+u_2})$&
$D(r_i, c, p^{u_i})$ exists,\\
&&&&  $i=1, 2$, $l<r_2$\\
\hline
\end{tabular*}
\tabnotetext[*]{tt9}{$p$ is any prime number.}
\end{table}

\begin{table}[b]
\tabcolsep=0pt
\caption{Comparisons among the $\mathit{NOA}(A_1, A; \rho_1, \rho_2)$'s
constructed by QTW, QAW, QA
and SLQ}
\label{comparison2}
\begin{tabular*}{\textwidth}{@{\extracolsep{\fill}}lcccc@{}}
\hline
\multicolumn{2}{@{}l}{\textbf{Methods}} & $\bolds{\rho_1(A_1)}$ & $\bolds{A}$ & \textbf{Constraints\tabnoteref{tt10}}\\
\hline
\multicolumn{2}{@{}l}{QTW} &$\mathit{OA}(p^{ku_1}, \frac{p^{ku_1}-1}{p^{u_1}-1}, p^{u_1}, 2)$
&$\mathit{OA}(p^{ku_2}, \frac{p^{ku_1}-1}{p^{u_1}-1}, p^{u_2}, 2)$&$2u_1\leq u_2+1$\\[3pt]
\multicolumn{2}{@{}l}{QAW} &$\mathit{OA}(s^2_1, 3, s_1, 2)$
&$\mathit{OA}(s^2_2, 3, s_2, 2)$&$ s_1<s_2, s_1|s_2$\\[3pt]
{QA}& I &$\mathit{OA}(p^{ku_1}, \frac{p^{ku_1}-1}{p^{u_1}-1},
p^{u_1}, 2)$
&$\mathit{OA}(p^{ku_2}, \frac{p^{ku_1}-1}{p^{u_1}-1}, p^{u_2}, 2)$& $ u_1<u_2,
u_1|u_2$\\
& II &$\mathit{OA}(p^{ku_1}, p^{u_1}+1, p^{u_1}, k)$
&$\mathit{OA}(p^{ku_2}, p^{u_1}+1, p^{u_2}, k)$ & $ u_1<u_2$,
$u_1|u_2$,\\
&&&& $p^{u_1}\geq k-1$ \\[3pt]
SLQ &Theorem~\ref{mthm1}&$\mathit{OA}(lp^{ku_1}, \frac
{p^{k}-1}{p-1}, p^{u_1}, 2)$
&$\mathit{OA}(p^{ku_2}, \frac{p^{k}-1}{p-1}, p^{u_2}, 2)$&$l<p^{k(u_2-u_1)}$,\\
&&&&
$u_1<u_2$ \\
&Theorem~\ref{mthm2}&$\mathit{OA}(lp^{ku_1}, \frac{p^{ku_1}-1}{p^{u_1}-1},
p^{u_1}, 2)$
&$\mathit{OA}(p^{ku_2}, \frac{p^{ku_1}-1}{p^{u_1}-1}, p^{u_2}, 2)$&$l<p^{k(u_2-u_1)}$,\\
&&&& $ u_1<u_2, u_1|u_2$\\
&Theorem~\ref{bush}&$\mathit{OA}(lp^{ku_1}, p^{u_1}+1, p^{u_1}, k)$
&$\mathit{OA}(p^{ku_2}, p^{u_1}+1, p^{u_2}, k)$ & $l<p^{k(u_2-u_1)}, u_1<u_2$, \\
&&&&$u_1|u_2$, $p^{u_1}\geq k-1$ \\
&Theorem~\ref{thml}& $\mathit{OA}(ln_1, m, p^{u_1}, t)$&$\mathit{OA}(n_1n_2, m,
p^{u_1+u_2}, t)$ &
$\mathit{OA}(n_i, m, p^{u_i}, t)$ exists,\\
&&&& $i=1,2$, $l<n_2$
\\
\hline
\end{tabular*}
\tabnotetext[*]{tt10}{$p$ is any prime number.}
\end{table}

\begin{table}
\caption{The $\mathit{NDM}(D_1, \ldots, D_I; \rho_1, \ldots, \rho_I)$'s constructed
in this paper for $I>2$}
\label{SNDMS}
\begin{tabular*}{\textwidth}{@{\extracolsep{\fill}}lcc@{}}
\hline
\textbf{Methods} & \multicolumn{1}{c}{$\bolds{\rho_i(D_i),\ i=1, \ldots, I}$} & \multicolumn{1}{c@{}}{\textbf{Constraints\tabnoteref{tt11}}} \\
\hline
Theorem~\ref{mthm3} & $D(p^{u_i}, p^{u_1}, p^{u_i})$ & $u_i<u_{i+1}$,
$u_i|u_{i+1}, i=1, \ldots, I-1$ \\
Theorem~\ref{thmcnd} & $D(\prod^i_{l=1}r_l, c, p^{\sum^i_{l=1}u_l})$ &
$D(r_i, c, p^{u_i})$ exists, $i=1, \ldots, I$ \\
\hline
\end{tabular*}
\tabnotetext[*]{tt11}{$p$ is any prime number.}
\end{table}
\begin{table}[b]
\caption{The $\mathit{NOA}(A_1, \ldots, A_I; \rho_1, \ldots, \rho_I)$'s constructed
in this paper for $I>2$}
\label{SNOAS}
\begin{tabular*}{\textwidth}{@{\extracolsep{\fill}}lcc@{}}
\hline
\textbf{Methods} & \multicolumn{1}{c}{$\bolds{\rho_i(A_i),\ i=1, \ldots, I}$} & \multicolumn{1}{c@{}}{\textbf{Constraints\tabnoteref{tt12}}} \\
\hline
Theorem \ref{mthm1} & $\mathit{OA}(p^{ku_i}, \frac{p^{k}-1}{p-1}, p^{u_i}, 2)$
        & $u_i<u_{i+1}, i=1, \ldots, I-1$\\
Theorem~\ref{mthm2} & $\mathit{OA}(p^{ku_i}, \frac{p^{ku_1}-1}{p^{u_1}-1},
p^{u_i}, 2)$
& $u_i<u_{i+1}, u_i|u_{i+1}, i=1, \ldots, I-1$ \\
Theorem~\ref{bush}&$\mathit{OA}(p^{ku_i}, p^{u_1}+1, p^{u_i}, k)$ & $p^{u_1}\geq k-1$, $u_i<u_{i+1}$, \\
&& $u_i|u_{i+1}, i=1, \ldots, I-1$ \\
Corollary~\ref{coroc2} & $\mathit{OA}(\prod^i_{l=1}n_l, m, p^{\sum^i_{l=1}u_l}, t)$
& $\mathit{OA}(n_i, m, p^{u_i}, t)$ exists, $i=1, \ldots, I$ \\
\hline
\end{tabular*}
\tabnotetext[*]{tt12}{$p$ is any prime number.}
\end{table}

From these tables and our construction methods, we can see that:
\begin{longlist}[(iii)]
\item[{(i)}] The proposed methods have more flexible choices of
the parameters, and thus can generate much more new NDMs and NOAs,
hence much more new NSFDs.

\item[{(ii)}] For NSFDs with two layers, some of the construction
results of QTW, QAW and QA can also be obtained by the proposed
methods. For example, in Table~\ref{comparison1}, by taking $l=1, p=2,
3, u_1=m, u_2=2, r_1=p^m$, and $r_2=c=p^2$, then the NDMs obtained by
our Theorem~\ref{thmcnd} are just those constructed by II of QAW. In
addition, most of the NOAs and NDMs obtained by the proposed methods
have no overlap with that of QTW, QAW and QA.

\item[{(iii)}] The proposed methods can generate various NDMs and
NOAs with more than two layers; see Tables~\ref{SNDMS} and \ref{SNOAS}.

\item[{(iv)}] Moreover, the methods for obtaining NOAs can also be
used to generate SOAs after some suitable modifications, which are
useful for constructing SSFDs for computer experiments with both
qualitative and quantitative factors [\citet{QW09}].
\end{longlist}

The newly proposed methods are easy to implement. The generated NSFDs
and SSFDs can be used not only in computer experiments, but also in
many other fields as mentioned in Section~\ref{sec1}.

\section*{Acknowledgments}
The authors thank the Editor, the Associate Editor and two referees
for their comments, which have led to improvements in the paper.




\printaddresses

\begin{thebibliography}{28}

\bibitem[\protect\citeauthoryear{Bose and Bush}{1952}]{BB2}
\begin{barticle}[mr]
\bauthor{\bsnm{Bose},~\bfnm{R.~C.}\binits{R.~C.}} \AND
\bauthor{\bsnm{Bush},~\bfnm{K.~A.}\binits{K.~A.}}
(\byear{1952}).
\btitle{Orthogonal arrays of strength two and three}.
\bjournal{Ann. Math. Statistics}
\bvolume{23}
\bpages{508--524}.
\bid{issn={0003-4851}, mr={0051204}}
\end{barticle}
\bptok{imsref}%
\endbibitem

\bibitem[\protect\citeauthoryear{Choi et~al.}{2008}]{CAK}
\begin{barticle}[auto:STB|2014/05/26|13:19:10]
\bauthor{\bsnm{Choi},~\bfnm{S.}\binits{S.}},
\bauthor{\bsnm{Alonso},~\bfnm{J.~J.}\binits{J.~J.}},
\bauthor{\bsnm{Kroo},~\bfnm{I.~M.}\binits{I.~M.}} \AND
\bauthor{\bsnm{Wintzer},~\bfnm{M.}\binits{M.}}
(\byear{2008}).
\btitle{Multifidelity design optimization of low-boom supersonic jets}.
\bjournal{Journal of Aircraft}
\bvolume{45}
\bpages{106--118}.
\end{barticle}
\bptok{imsref}%
\endbibitem

\bibitem[\protect\citeauthoryear{Dewettinck et~al.}{1999}]{DVD}
\begin{barticle}[auto:STB|2014/05/26|13:19:10]
\bauthor{\bsnm{Dewettinck},~\bfnm{K.}\binits{K.}},
\bauthor{\bsnm{Visscher},~\bfnm{A.~D.}\binits{A.~D.}},
\bauthor{\bsnm{Deroo},~\bfnm{L.}\binits{L.}} \AND
\bauthor{\bsnm{Huyghebaert},~\bfnm{A.}\binits{A.}}
(\byear{1999}).
\btitle{Modeling the steady-state thermodynamic operation point of top-spray fluidized bed processing}.
\bjournal{Journal of Food Engineering}
\bvolume{39}
\bpages{131--143}.
\end{barticle}
\bptok{imsref}%
\endbibitem

\bibitem[\protect\citeauthoryear{Fang, Li and Sudjianto}{2006}]{FLS}
\begin{bbook}[mr]
\bauthor{\bsnm{Fang},~\bfnm{Kai-Tai}\binits{K.-T.}},
\bauthor{\bsnm{Li},~\bfnm{Runze}\binits{R.}} \AND
\bauthor{\bsnm{Sudjianto},~\bfnm{Agus}\binits{A.}}
(\byear{2006}).
\btitle{Design and Modeling for Computer Experiments}.
\bpublisher{Chapman \& Hall/CRC},
\blocation{Boca Raton, FL}.
\bid{mr={2223960}}
\bptnote{check year}%
\end{bbook}
\bptok{imsref}%
\endbibitem

\bibitem[\protect\citeauthoryear{Fasshauer}{2007}]{F07}
\begin{bbook}[mr]
\bauthor{\bsnm{Fasshauer},~\bfnm{Gregory~E.}\binits{G.~E.}}
(\byear{2007}).
\btitle{Meshfree Approximation Methods with {MATLAB}}.
\bseries{Interdisciplinary Mathematical Sciences}
\bvolume{6}.
\bpublisher{World Scientific},
\blocation{Hackensack, NJ}.
\bid{mr={2357267}}
\end{bbook}
\bptok{imsref}%
\endbibitem

\bibitem[\protect\citeauthoryear{Floater and Iske}{1996}]{FI}
\begin{barticle}[mr]
\bauthor{\bsnm{Floater},~\bfnm{Michael~S.}\binits{M.~S.}} \AND
\bauthor{\bsnm{Iske},~\bfnm{Armin}\binits{A.}}
(\byear{1996}).
\btitle{Multistep scattered data interpolation using compactly supported radial basis functions}.
\bjournal{J. Comput. Appl. Math.}
\bvolume{73}
\bpages{65--78}.
\bid{doi={10.1016/0377-0427(96)00035-0}, issn={0377-0427}, mr={1424869}}
\end{barticle}
\bptok{imsref}%
\endbibitem

\bibitem[\protect\citeauthoryear{Haaland and Qian}{2010}]{HQ}
\begin{barticle}[mr]
\bauthor{\bsnm{Haaland},~\bfnm{Ben}\binits{B.}} \AND
\bauthor{\bsnm{Qian},~\bfnm{Peter~Z.~G.}\binits{P.~Z.~G.}}
(\byear{2010}).
\btitle{An approach to constructing nested space-filling designs for multi-fidelity computer experiments}.
\bjournal{Statist. Sinica}
\bvolume{20}
\bpages{1063--1075}.
\bid{issn={1017-0405}, mr={2729853}}
\end{barticle}
\bptok{imsref}%
\endbibitem

\bibitem[\protect\citeauthoryear{Haaland and Qian}{2011}]{HQ2}
\begin{barticle}[mr]
\bauthor{\bsnm{Haaland},~\bfnm{Ben}\binits{B.}} \AND
\bauthor{\bsnm{Qian},~\bfnm{Peter~Z.~G.}\binits{P.~Z.~G.}}
(\byear{2011}).
\btitle{Accurate emulators for large-scale computer experiments}.
\bjournal{Ann. Statist.}
\bvolume{39}
\bpages{2974--3002}.
\bid{doi={10.1214/11-AOS929}, issn={0090-5364}, mr={3012398}}
\end{barticle}
\bptok{imsref}%
\endbibitem

\bibitem[\protect\citeauthoryear{Han et~al.}{2009}]{HSNB}
\begin{barticle}[mr]
\bauthor{\bsnm{Han},~\bfnm{Gang}\binits{G.}},
\bauthor{\bsnm{Santner},~\bfnm{Thomas~J.}\binits{T.~J.}},
\bauthor{\bsnm{Notz},~\bfnm{William~I.}\binits{W.~I.}} \AND
\bauthor{\bsnm{Bartel},~\bfnm{Donald~L.}\binits{D.~L.}}
(\byear{2009}).
\btitle{Prediction for computer experiments having quantitative and qualitative input variables}.
\bjournal{Technometrics}
\bvolume{51}
\bpages{278--288}.
\bid{doi={10.1198/tech.2009.07132}, issn={0040-1706}, mr={2751072}}
\end{barticle}
\bptok{imsref}%
\endbibitem

\bibitem[\protect\citeauthoryear{Hedayat, Sloane and Stufken}{1999}]{HSS}
\begin{bbook}[mr]
\bauthor{\bsnm{Hedayat},~\bfnm{A.~S.}\binits{A.~S.}},
\bauthor{\bsnm{Sloane},~\bfnm{N.~J.~A.}\binits{N.~J.~A.}} \AND
\bauthor{\bsnm{Stufken},~\bfnm{John}\binits{J.}}
(\byear{1999}).
\btitle{Orthogonal Arrays: Theory and Applications}.
\bpublisher{Springer},
\blocation{New York}.
\bid{doi={10.1007/978-1-4612-1478-6}, mr={1693498}}
\end{bbook}
\bptok{imsref}%
\endbibitem

\bibitem[\protect\citeauthoryear{Herstein}{1996}]{H96}
\begin{bbook}[mr]
\bauthor{\bsnm{Herstein},~\bfnm{I.~N.}\binits{I.~N.}}
(\byear{1996}).
\btitle{Abstract Algebra},
\bedition{3rd} ed.
\bpublisher{Prentice Hall},
\blocation{Upper Saddle River, NJ}.
\bid{mr={1375019}}
\end{bbook}
\bptok{imsref}%
\endbibitem

\bibitem[\protect\citeauthoryear{Husslage et~al.}{2003}]{HDHSS}
\begin{barticle}[auto:STB|2014/05/26|13:19:10]
\bauthor{\bsnm{Husslage},~\bfnm{B.}\binits{B.}},
\bauthor{\bsnm{Dam},~\bfnm{E.~V.}\binits{E.~V.}},
\bauthor{\bsnm{Hertog},~\bfnm{D.~D.}\binits{D.~D.}},
\bauthor{\bsnm{Stehouwer},~\bfnm{P.}\binits{P.}} \AND
\bauthor{\bsnm{Stinstra},~\bfnm{E.}\binits{E.}}
(\byear{2003}).
\btitle{Collaborative metamodeling:
Coordinating simulation-based product design}.
\bjournal{Concurrent Eng.}
\bvolume{11}
\bpages{267--278}.
\end{barticle}
\bptok{imsref}%
\endbibitem

\bibitem[\protect\citeauthoryear{McKay, Beckman and Conover}{1979}]{MBC}
\begin{barticle}[mr]
\bauthor{\bsnm{McKay},~\bfnm{M.~D.}\binits{M.~D.}},
\bauthor{\bsnm{Beckman},~\bfnm{R.~J.}\binits{R.~J.}} \AND
\bauthor{\bsnm{Conover},~\bfnm{W.~J.}\binits{W.~J.}}
(\byear{1979}).
\btitle{A comparison of three methods for selecting values of input variables in the analysis of output from a computer code}.
\bjournal{Technometrics}
\bvolume{21}
\bpages{239--245}.
\bid{doi={10.2307/1268522}, issn={0040-1706}, mr={0533252}}
\end{barticle}
\bptok{imsref}%
\endbibitem

\bibitem[\protect\citeauthoryear{Molina-Crist\'{o}bal et al.}{2010}]{MPS}
\begin{bmisc}[auto:STB|2014/05/26|13:19:10]
\bauthor{\bsnm{Molina-Crist\'{o}bal},~\bfnm{A.}\binits{A.}},
\bauthor{\bsnm{Palmer},~\bfnm{P.~R.}\binits{P.~R.}},
\bauthor{\bsnm{Skinner},~\bfnm{B.~A.}\binits{B.~A.}}
\AND
\bauthor{\bsnm{Parks},~\bfnm{G.~T.}\binits{G.~T.}}
(\byear{2010}).
\bhowpublished{Multi-fidelity simulation modelling in optimization of a
submarine propulsion system.
In \textit{Proceedings of the 2010 IEEE Vehicle Power and Propulsion
Conference (VPPC)}.
Lille, France}.
\end{bmisc}
\bptok{imsref}%
\endbibitem

\bibitem[\protect\citeauthoryear{Mukerjee, Qian and Jeff~Wu}{2008}]{MQW}
\begin{barticle}[mr]
\bauthor{\bsnm{Mukerjee},~\bfnm{Rahul}\binits{R.}},
\bauthor{\bsnm{Qian},~\bfnm{Peter~Z.~G.}\binits{P.~Z.~G.}} \AND
\bauthor{\bsnm{Jeff Wu},~\bfnm{C.~F.}\binits{C.~F.}}
(\byear{2008}).
\btitle{On the existence of nested orthogonal arrays}.
\bjournal{Discrete Math.}
\bvolume{308}
\bpages{4635--4642}.
\bid{doi={10.1016/j.disc.2007.08.096}, issn={0012-365X}, mr={2438169}}
\end{barticle}
\bptok{imsref}%
\endbibitem

\bibitem[\protect\citeauthoryear{Qian}{2009}]{Q09}
\begin{barticle}[mr]
\bauthor{\bsnm{Qian},~\bfnm{Peter~Z.~G.}\binits{P.~Z.~G.}}
(\byear{2009}).
\btitle{Nested {L}atin hypercube designs}.
\bjournal{Biometrika}
\bvolume{96}
\bpages{957--970}.
\bid{doi={10.1093/biomet/asp045}, issn={0006-3444}, mr={2767281}}
\end{barticle}
\bptok{imsref}%
\endbibitem

\bibitem[\protect\citeauthoryear{Qian}{2012}]{Q12}
\begin{barticle}[mr]
\bauthor{\bsnm{Qian},~\bfnm{Peter~Z.~G.}\binits{P.~Z.~G.}}
(\byear{2012}).
\btitle{Sliced {L}atin hypercube designs}.
\bjournal{J. Amer. Statist. Assoc.}
\bvolume{107}
\bpages{393--399}.
\bid{doi={10.1080/01621459.2011.644132}, issn={0162-1459}, mr={2949368}}
\end{barticle}
\bptok{imsref}%
\endbibitem

\bibitem[\protect\citeauthoryear{Qian and Ai}{2010}]{QA}
\begin{barticle}[mr]
\bauthor{\bsnm{Qian},~\bfnm{Peter~Z.~G.}\binits{P.~Z.~G.}} \AND
\bauthor{\bsnm{Ai},~\bfnm{Mingyao}\binits{M.}}
(\byear{2010}).
\btitle{Nested lattice sampling: A new sampling scheme derived by randomizing nested orthogonal arrays}.
\bjournal{J. Amer. Statist. Assoc.}
\bvolume{105}
\bpages{1147--1155}.
\bid{doi={10.1198/jasa.2010.tm09365}, issn={0162-1459}, mr={2752610}}
\end{barticle}
\bptok{imsref}%
\endbibitem

\bibitem[\protect\citeauthoryear{Qian, Ai and Wu}{2009}]{QAW}
\begin{barticle}[mr]
\bauthor{\bsnm{Qian},~\bfnm{Peter~Z.~G.}\binits{P.~Z.~G.}},
\bauthor{\bsnm{Ai},~\bfnm{Mingyao}\binits{M.}} \AND
\bauthor{\bsnm{Wu},~\bfnm{C.~F.~Jeff}\binits{C.~F.~J.}}
(\byear{2009}).
\btitle{Construction of nested space-filling designs}.
\bjournal{Ann. Statist.}
\bvolume{37}
\bpages{3616--3643}.
\bid{doi={10.1214/09-AOS690}, issn={0090-5364}, mr={2549572}}
\end{barticle}
\bptok{imsref}%
\endbibitem

\bibitem[\protect\citeauthoryear{Qian, Tang and Wu}{2009}]{QTW}
\begin{barticle}[mr]
\bauthor{\bsnm{Qian},~\bfnm{Peter~Z.~G.}\binits{P.~Z.~G.}},
\bauthor{\bsnm{Tang},~\bfnm{Boxin}\binits{B.}} \AND
\bauthor{\bsnm{Wu},~\bfnm{C.~F.~Jeff}\binits{C.~F.~J.}}
(\byear{2009}).
\btitle{Nested space-filling designs for computer experiments with two levels of accuracy}.
\bjournal{Statist. Sinica}
\bvolume{19}
\bpages{287--300}.
\bid{issn={1017-0405}, mr={2487890}}
\end{barticle}
\bptok{imsref}%
\endbibitem

\bibitem[\protect\citeauthoryear{Qian and Wu}{2009}]{QW09}
\begin{barticle}[mr]
\bauthor{\bsnm{Qian},~\bfnm{Peter~Z.~G.}\binits{P.~Z.~G.}} \AND
\bauthor{\bsnm{Wu},~\bfnm{C.~F.~Jeff}\binits{C.~F.~J.}}
(\byear{2009}).
\btitle{Sliced space-filling designs}.
\bjournal{Biometrika}
\bvolume{96}
\bpages{945--956}.
\bid{doi={10.1093/biomet/asp044}, issn={0006-3444}, mr={2767280}}
\end{barticle}
\bptok{imsref}%
\endbibitem

\bibitem[\protect\citeauthoryear{Qian, Wu and Wu}{2008}]{QWW}
\begin{barticle}[mr]
\bauthor{\bsnm{Qian},~\bfnm{Peter~Z.~G.}\binits{P.~Z.~G.}},
\bauthor{\bsnm{Wu},~\bfnm{Huaiqing}\binits{H.}} \AND
\bauthor{\bsnm{Wu},~\bfnm{C.~F.~Jeff}\binits{C.~F.~J.}}
(\byear{2008}).
\btitle{Gaussian process models for computer experiments with qualitative and quantitative factors}.
\bjournal{Technometrics}
\bvolume{50}
\bpages{383--396}.
\bid{doi={10.1198/004017008000000262}, issn={0040-1706}, mr={2457574}}
\end{barticle}
\bptok{imsref}%
\endbibitem

\bibitem[\protect\citeauthoryear{Santner, Williams and Notz}{2003}]{SWN}
\begin{bbook}[mr]
\bauthor{\bsnm{Santner},~\bfnm{Thomas~J.}\binits{T.~J.}},
\bauthor{\bsnm{Williams},~\bfnm{Brian~J.}\binits{B.~J.}} \AND
\bauthor{\bsnm{Notz},~\bfnm{William~I.}\binits{W.~I.}}
(\byear{2003}).
\btitle{The Design and Analysis of Computer Experiments}.
\bpublisher{Springer},
\blocation{New York}.
\bid{doi={10.1007/978-1-4757-3799-8}, mr={2160708}}
\end{bbook}
\bptok{imsref}%
\endbibitem

\bibitem[\protect\citeauthoryear{Schmidt, Cruz and Iyengar}{2005}]{SCI}
\begin{barticle}[auto:STB|2014/05/26|13:19:10]
\bauthor{\bsnm{Schmidt},~\bfnm{R.~R.}\binits{R.~R.}},
\bauthor{\bsnm{Cruz},~\bfnm{E.~E.}\binits{E.~E.}} \AND
\bauthor{\bsnm{Iyengar},~\bfnm{M.~K.}\binits{M.~K.}}
(\byear{2005}).
\btitle{Challenges of data center thermal management}.
\bjournal{IBM Journal of Research and Development}
\bvolume{49}
\bpages{709--723}.
\end{barticle}
\bptok{imsref}%
\endbibitem

\bibitem[\protect\citeauthoryear{Tang}{1993}]{T93}
\begin{barticle}[mr]
\bauthor{\bsnm{Tang},~\bfnm{Boxin}\binits{B.}}
(\byear{1993}).
\btitle{Orthogonal array-based {L}atin hypercubes}.
\bjournal{J. Amer. Statist. Assoc.}
\bvolume{88}
\bpages{1392--1397}.
\bid{issn={0162-1459}, mr={1245375}}
\end{barticle}
\bptok{imsref}%
\endbibitem

\bibitem[\protect\citeauthoryear{Williams, Morris and Santner}{2009}]{WMS}
\begin{bmisc}[auto:STB|2014/05/26|13:19:10]
\bauthor{\bsnm{Williams},~\bfnm{B.}\binits{B.}},
\bauthor{\bsnm{Morris},~\bfnm{M.}\binits{M.}} \AND
\bauthor{\bsnm{Santner},~\bfnm{T.}\binits{T.}}
(\byear{2009}).
\bhowpublished{Using multiple computer models/multiple
data sources simultaneously to infer calibration parameters.
Paper presented at the 2009 INFORMS Annual Conference, October 11--14,
San Diego, CA}.
\end{bmisc}
\bptok{imsref}%
\endbibitem

\bibitem[\protect\citeauthoryear{Xu, Haaland and Qian}{2011}]{XHQ}
\begin{barticle}[mr]
\bauthor{\bsnm{Xu},~\bfnm{Xu}\binits{X.}},
\bauthor{\bsnm{Haaland},~\bfnm{Ben}\binits{B.}} \AND
\bauthor{\bsnm{Qian},~\bfnm{Peter~Z.~G.}\binits{P.~Z.~G.}}
(\byear{2011}).
\btitle{Sudoku-based space-filling designs}.
\bjournal{Biometrika}
\bvolume{98}
\bpages{711--720}.
\bid{doi={10.1093/biomet/asr024}, issn={0006-3444}, mr={2836416}}
\end{barticle}
\bptok{imsref}%
\endbibitem

\bibitem[\protect\citeauthoryear{Zhou, Qian and Zhou}{2011}]{ZQZ}
\begin{barticle}[mr]
\bauthor{\bsnm{Zhou},~\bfnm{Qiang}\binits{Q.}},
\bauthor{\bsnm{Qian},~\bfnm{Peter~Z.~G.}\binits{P.~Z.~G.}} \AND
\bauthor{\bsnm{Zhou},~\bfnm{Shiyu}\binits{S.}}
(\byear{2011}).
\btitle{A simple approach to emulation for computer models with qualitative and quantitative factors}.
\bjournal{Technometrics}
\bvolume{53}
\bpages{266--273}.
\bid{doi={10.1198/TECH.2011.10025}, issn={0040-1706}, mr={2857704}}
\end{barticle}
\bptok{imsref}%
\endbibitem

\end{thebibliography}
\end{document}